\documentclass[a4paper,11pt]{article}
\pdfoutput=1

\usepackage{jheppub}
\usepackage{graphicx}
\usepackage{amsmath}
\usepackage{mathrsfs}
\usepackage[utf8]{inputenc}

\usepackage{marginnote}

\newcommand{\eq}[1]{eq.~\eqref{eq:#1}}
\newcommand{\eqs}[2]{eqs.~\eqref{eq:#1} and \eqref{eq:#2}}
\newcommand{\eqss}[2]{eqs.~\eqref{eq:#1} - \eqref{eq:#2}}
\renewcommand{\sec}[1]{section~\ref{sec:#1}}

\newcommand{\app}[1]{appendix~\ref{app:#1}}

\newcommand{\mycites}[1]{refs.~\cite{#1}}
\newcommand{\mycite}[1]{ref.~\cite{#1}}

\newcommand{\Tau}{\mathcal{T}}

\newcommand{\lqcd}{\Lambda_\mathrm{QCD}}

\allowdisplaybreaks[2]

\newcommand{\ord}[1]{{\mathcal O}(#1)}

\newcommand{\ORD}[1]{{\mathcal O}\biggl(#1\biggr)}

\newcommand{\Mae}[3]{\bigl\langle#1\bigr\rvert#2\bigr\rvert#3\bigr\rangle}

\newcommand{\bn}{\bar{n}}

\newcommand{\df}{\mathrm{d}}

\newcommand{\Li}{\mathrm{Li}}

\newcommand{\eps}{\epsilon}

\newcommand{\cB}{{\mathcal B}}
\newcommand{\cL}{{\mathcal L}}
\newcommand{\cI}{{\mathcal I}}

\newcommand{\nn}{\nonumber}

\newcommand{\hp}{\hat{p}}
\newcommand{\bnP}{\overline {\mathcal P}}

\newcommand{\conv}{\!\otimes\!}
\newcommand{\convz}{\!\otimes_z\!}

\newcommand{\zero}{{(0)}}
\newcommand{\one}{{(1)}}
\newcommand{\two}{{(2)}}

\newcommand{\ktsq}{\vec{k}_{\perp}^{\,2}}
\newcommand{\kt}{\vec{k}_{\perp}}

\newcommand{\msb}{\overline{\mathrm{MS}}}

\newcommand{\cusp}{\mathrm{cusp}}

\preprint{\begin{flushright}
		CERN-TH-2020-065\\
		FR-PHENO-2020-003
\end{flushright}}

\title{The Fully-Differential Gluon Beam Function at NNLO}

\author[a]{Jonathan R.~Gaunt}
\author[b]{and Maximilian Stahlhofen}

\affiliation[a]{CERN Theory Division, 1211 Geneva 23, Switzerland}
\affiliation[b]{Albert-Ludwigs-Universit\"at Freiburg, Physikalisches Institut, D-79104 Freiburg, Germany}

\emailAdd{jonathan.richard.gaunt@cern.ch}
\emailAdd{maximilian.stahlhofen@physik.uni-freiburg.de}

\abstract{
	The fully-differential beam function (dBF) is a universal ingredient in resummed predictions of hadron collider observables that probe the full kinematics of the incoming parton from each colliding proton -- the virtuality and transverse momentum as well as the light-cone momentum fraction $x$. In this paper we compute the matching coefficients between the unpolarized gluon dBF and the usual parton distribution functions (PDFs) at the two-loop order. For observables probing both the virtuality and transverse momentum of incoming gluons, our results provide the part of the NNLO singular cross section related to collinear initial-state radiation, and are required for the resummation of large logarithms through N$^3$LL. Further to this, the dBF is closely linked to the beam function appearing in a generalized version of threshold factorization, via a simple integration. By performing this integration for the two-loop gluon matching coefficients, we also obtain the corresponding quantities for the generalized threshold beam function.
}

\begin{document}
	
\maketitle

\section{Introduction}
\label{sec:intro}

Beam functions encode the effect of collinear initial state radiation (ISR) on the measurement of some observable $\Tau$ at a hadron collider~\cite{Stewart:2009yx, Stewart:2010qs}. If $\Tau$ is a perturbative scale ($\Tau \gg \Lambda_{\mathrm{QCD}}$), these functions can be expressed in terms of the convolution of the standard PDFs $f_j$ and perturbatively-calculable matching coefficients $\cI_{ij}$. 
The indices $i,j = \{q_i,\bar q_i, g\}$ denote the different types of partons in QCD. 
In this paper we focus on the beam functions $B_i(t, x, \kt)$~\cite{Jain:2011iu}%
\footnote{See also \mycite{Mantry:2009qz} for related beam functions defined in impact parameter space. }, which can be thought of as a parton density completely differential in all of the kinematic variables of the `active' parton that enters the hard process. It is a main ingredient in factorized cross sections for observables sensitive to $\kt$ and $t$ in the limit $Q^2 \gg t \sim \ktsq \gg \Lambda^2_{\mathrm{QCD}}$, where $Q$ represents the scale of the hard interaction. The light-cone coordinates%
\footnote{Here we use the SCET conventions for light-cone coordinates. 
	That is, for a Lorentz vector $v$ the components $(v^+, v^-, v_{\perp})$ are defined by $v^\mu = v^+ n^\mu/2 + v^- \bar{n}^\mu/2 + v_{\perp}^\mu$, where $n$ and $\bar{n}$ are oppositely-pointing lightcone vectors and $v_{\perp}^\mu$ is a vector transverse to these directions: $n \cdot \bar{n}$ = 2, $n^2 = \bar{n}^2 = 0$, $n \cdot v_{\perp} = \bar{n} \cdot v_{\perp} = 0$. In our notation $\vec{v}_{\perp}$ is a Euclidean transverse vector: $\vec{v}_{\perp}^2 = -v_{\perp}^2$.}
of the active parton are given in terms of the arguments of this beam function by $(xP^-, -t/[xP^-], \kt)$, with $P^-$ the large component of the proton momentum. The $B_i(t, x, \kt)$ are often referred to as fully-differential beam functions (dBFs) \cite{Jain:2011iu}, or fully unintegrated PDFs \cite{Collins:2007ph,Rogers:2008jk}. 

The full set of dBF matching coefficients $\cI_{ij}$ was calculated at one loop in \mycite{Jain:2011iu}. In \mycite{Gaunt:2014xxa}, we computed the quark matching coefficients $\cI_{qj}$ at the two-loop level. The goal of the present paper is to extend this computation to the case of the gluon matching coefficients $\cI_{gj}$. In an unpolarized proton, there are two varieties of gluon dBFs: the unpolarized and linearly polarized dBFs (similar to what is observed for the transverse momentum dependent beam functions, or TMD PDFs \cite{Mulders:2000sh}). 
We limit our attention here to the computation of the former quantity at two loops, which is sufficient for NNLO  or N$^3$LL resummed predictions of the most important  gluon-initiated processes like Higgs production, see \sec{defsandprops}.
 
Our motivation for performing this calculation is the large and growing list of potential applications for the dBF: 
\begin{itemize}
\item{The gluon dBF is required to make precise resummed predictions of the Higgs transverse momentum distribution in the presence of a jet veto, where the jet veto is imposed via a cut on a virtuality-sensitive observable \cite{Jain:2011iu, Tackmann:2012bt, Procura:2014cba, Gangal:2014qda}. In Higgs measurements, veto constraints of this kind are often imposed to enhance the Higgs signal over the backgrounds \cite{Aaboud:2018jqu, Sirunyan:2018egh}, or categorise the cross section in terms of different initial states \cite{Aaboud:2018xdt}. }
\item{Beam functions are a core component of the {\sc Geneva} methodology for combining parton showers with higher order resummation and fixed order computations \cite{Alioli:2012fc,Alioli:2013hqa,Alioli:2015toa, Alioli:2019qzz}.
	The two-loop gluon dBF would be a necessary ingredient in constructing a shower in the context of (for example) Higgs production that was simultaneously NNLL$'$ accurate in both the Higgs $p_T$ and beam thrust $\Tau_0$, and matched to NNLO.}
\item{The two-loop gluon dBF could be utilised in multi-differential extensions of the $N$-jettiness subtraction scheme \cite{Gaunt:2015pea, Boughezal:2015dva} in computations of NNLO gluon-initiated processes. This is discussed in section 3.4 of \mycite{Gaunt:2015pea}.
}
\item{Finally, in \mycite{Lustermans:2019cau} a generalized threshold factorization theorem was derived for colour singlet production that holds when the momentum fraction of one of the incoming partons approaches unity. This corresponds to the limit of large rapidity but generic invariant mass of the produced colour singlet. A beam function $\tilde{B}_j(\tilde{t},x)$ appears in this factorization theorem that can be obtained from the dBF as follows:
\begin{align} \label{eq:dBF2genBF}
\tilde{B}_j(\tilde{t},x) = \int \mathrm{d}^2 \kt B_j\biggl(\tilde{t} - \dfrac{\ktsq}{2}, x, \kt \biggr)\,.
\end{align}
Thus one can straightforwardly obtain the two-loop matching coefficients for the generalized threshold beam function from our two-loop results for the gluon dBF. In this paper we actually perform this exercise, and include the results below.
Note that only the unpolarized part of the dBF contributes in \eq{dBF2genBF}.
}
\end{itemize}

The paper is organised as follows. In \sec{defsandprops} we give the operator definition of the dBFs and recall their key properties as well as their distributional structure. For most of the calculational details we refer to previous work. Section \ref{sec:results} contains the novel results for the two-loop gluon dBF matching coefficients. In \sec{genBF} (and \app{genBFapp}) we present the corresponding results for the generalized threshold beam function. Finally we conclude in section \ref{sec:conclusions}.

\section{Definitions and Properties of the dBF}
\label{sec:defsandprops}

Our calculation of the two-loop gluon dBF is performed in close analogy to our calculation of the  two-loop quark dBF~\cite{Gaunt:2014xxa} and largely based on the two-loop calculation of the virtuality-dependent beam functions~\cite{Gaunt:2014xga,Gaunt:2014cfa}. In particular the integrand of the NNLO gluon beam function is taken from \mycite{Gaunt:2014cfa}. The (loop) integrations are then performed with the additional constraint on the transverse momentum crossing the unitarity cut as in \mycite{Gaunt:2014xxa}.
For details on the calculational methods we therefore refer the reader to the above three references and shall only review the generic structure and some important properties of the dBFs here.

The bare gluon dBF is defined by the operator matrix element~\cite{Jain:2011iu}
\begin{align} \label{eq:dBg_def}
&B_g^{\mu\nu}(t, x, \kt) =  \nn\\
&-k^- \theta(k^-)\Mae{p_n(p^-)}{\cB_{n\perp}^{\mu c}(0)\, \delta(t - k^-\hp^+)  
	\bigl[\delta(k^- \!\!- \bnP_n) \delta^\two(\kt \!- \vec{\mathcal{P}}_{n\perp}) 
	\cB_{n\perp}^{\nu c}(0)
	 \bigr]}{p_n(p^-)}
 \,,
\end{align}
in soft-collinear effective theory (SCET)~\cite{Bauer:2000ew, Bauer:2000yr, Bauer:2001ct, Bauer:2001yt, Bauer:2002nz, Beneke:2002ph}.
Here $p_n(p^-)$ denotes the incoming spin-averaged proton state with lightlike momentum $p^\mu=p^- n^\mu/2$, $x\equiv k^-/p^-$ and $\cB_{n\perp}^\mu$ is the gauge-invariant gluon field strength operator in SCET.
The measurement delta functions involve the SCET label momentum operators $\vec{\mathcal{P}}_{n\perp}$ and $\bnP_n$~\cite{Bauer:2001ct} as well as the usual (derivative) plus-momentum operator $\hp^+$. The label momentum operators act to the right on the SCET fields in $\cB_{n\perp}^\mu$, while $\hp^+$ also acts on the external proton state.
 For further details and explanations on the relevant SCET operators, see e.g.\ \mycites{Stewart:2009yx, Stewart:2010qs, Gaunt:2014xga}.

The gluon dBF is a Lorentz tensor transverse to the lightcone directions $n^\mu$ and $\bn^\mu$, i.e.\ $n_\mu B_g^{\mu\nu} = \bn_\mu B_g^{\mu\nu} = 0$, as a consequence of the transverse polarization of the physical gluon field operators generating it. As it depends on the external two-dimensional transverse momentum vector $\kt$ the renormalized gluon dBF (with renormalization scale $\mu$) naturally decomposes into two orthogonal tensor structures%
\footnote{Orthogonal in the sense that the trace of the contracted product of the two tensors vanishes.}
 in four dimensions:
\begin{align} \label{eq:dBg_tensor}
&B_g^{\mu\nu}(t, x, \kt,\mu) =
B_g(t, x, \ktsq,\mu) \frac{g^{\mu \nu}_\perp}{2} 
+ \mathscr{B}_g(t, x, \ktsq,\mu) 
\Big( \frac{g^{\mu \nu}_\perp}{2}  + \frac{k^\mu_\perp k^\nu_\perp}{\ktsq} \Big)
\,.
\end{align}
The two scalar coefficients $B_g$ and $\mathscr{B}_g$ depend on $\kt$ only through $\ktsq$ and can be regarded as independent (unpolarized and linearly polarized) beam functions. In particular, the corresponding projections of the dBF operator do not mix under renormalization. We can thus choose different regularization schemes for the computation of the respective bare functions.
While a momentum space calculation of $\mathscr{B}_g$ requires a regularization scheme that treats $\kt$ as strictly two-dimensional vector, there is in fact no such restriction for $B_g$.
That is because we may replace
\begin{align}
\label{eq:deltaReplace}
\delta^{(2)}(\kt-\mathcal{P}_\perp) \,\rightarrow\, \frac{1}{\pi}\delta(\ktsq-\mathcal{P}_\perp^2)\,,
\end{align}
in the dBF matrix element $B_g \equiv B_{g\mu}^\mu$, \eq{dBg_def}, which is not possible for $\mathscr{B}_g$ since the second term in \eq{dBg_tensor} depends on $\kt$ rather than $\ktsq$.
We can thus employ conventional dimensional regularization (CDR) for the computation of $B_g$ by interpreting $\kt$ on the RHS of \eq{deltaReplace} as $(d-2)$ dimensional vector with $d=4-2\eps$. 
This in turn is equivalent to replacing
\begin{align}
\label{eq:deltaReplaceCDR}
 \frac{1}{\pi}\delta(\ktsq-\mathcal{P}_\perp^2) \,\rightarrow\,
 \frac{(\ktsq)^{-\epsilon}}{\Gamma(1-\epsilon)\pi^{\epsilon}}\; \delta^{(d-2)}(\kt-\mathcal{P}_\perp)
\end{align}
in our calculation of the bare $B_g$ and in analogy to what we did in our calculation of the two-loop quark dBF~\cite{Gaunt:2014xxa}.
While intermediate (bare) results may differ depending on the dimensional regularization scheme, the renormalized results do not.
The argument closely follows the one given in \mycite{Luebbert:2016itl} in the context of the 
transverse momentum dependent soft function.
It is based on the observation that all $1/\eps^n$ poles associated with ultraviolet, infrared, or rapidity divergences are, after the subtraction of subdivergences in the renormalization or matching procedure, proportional to $\delta(\ktsq)=\pi \delta^\two(\kt)$ and therefore cannot promote the $\ord{\eps}$ differences in the transverse momentum measurement to a finite contribution.
Throughout this work we use the $\msb$ renormalization scheme.

For all gluon-initiated processes that are insensitive to the polarization of the incoming gluons, like Higgs or (azimuthally averaged) top pair production (cf.\ \mycites{Zhu:2012ts,Li:2013mia})
 only the NLO result for $\mathscr{B}_g$~\cite{Jain:2011iu} contributes to NNLO (leading power) or N$^3$LL resummed predictions.
On the other hand this precision does require the NNLO result of $B_g$.
The reason is that due to the orthogonality of the two tensor structures in \eq{dBg_tensor} only diagonal terms $\propto B_g \otimes B_g$ or $\propto \mathscr{B}_g \otimes \mathscr{B}_g$ and no mixed terms occur in the factorized cross section for these processes.
At LO (tree-level) $ \mathscr{B}_g$ vanishes (unlike $B_g$)  and thus only the NLO $\otimes$ NLO term of $\mathscr{B}_g \otimes \mathscr{B}_g$ type survives at NNLO.

It is the aim of this paper to compute the so far unknown NNLO correction to $B_g$, i.e.\ more precisely the matching coefficient $\cI_{gj}(t,z,\ktsq,\mu)$ in the OPE~\cite{Fleming:2006cd, Stewart:2009yx}
\begin{align} \label{eq:dBi_OPE}
B_g(t,x,\ktsq,\mu) 
& = \sum_j \int_x^1 \frac{\df z}{z}
\cI_{gj}\Big(t,z,\ktsq,\mu\Big) f_j(\frac{x}{z},\mu)
\bigg[1+ \ORD{\frac{\lqcd^2}{t},\frac{\lqcd^2}{\ktsq}}\bigg]\,.
\end{align}
In practice we determine the $\cI_{gj}$ coefficients from a perturbative calculation along the lines of \mycite{Gaunt:2014xxa}, where the incoming proton is replaced by a single parton on both sides of \eq{dBi_OPE}.
The relevant two-loop diagrams (in lightcone axial gauge) are the same as the ones shown in \mycite{Gaunt:2014cfa}.
Note that $\ktsq$ is constrained by~\cite{Gaunt:2014xxa}
\begin{align} \label{eq:ktsqlimits}
\frac{1-z}{z} \, t \,\ge\, \ktsq \,\ge\,0\,,
\end{align}
where $z$ is the large lightcone momentum fraction argument of $\cI_{gj}$ as in \eq{dBi_OPE}.
That is because  the total invariant mass of the collinear radiation emitted in the transition of parton $j$ to the active gluon must be non-negative. As $z\ge x \ge 0$, \eq{ktsqlimits} also holds for $z \to x$.
Integrating over the full range of $\ktsq$ yields the virtuality ($t$) dependent gluon beam function. For the corresponding matching coefficients this means
\begin{align}
\label{eq:I_int}
\int \!\! \df^2k_\perp \, \cI_{gj}(t,z,\ktsq,\mu) = \pi \!\int \!\! \df \big(\ktsq\big) \, \cI_{gj}(t,z,\ktsq,\mu) = \cI_{gj}(t,z, \mu)\,,
\end{align}
with the $\cI_{gj}(t,x,\mu)$ known to NNLO~\cite{Berger:2010xi,Gaunt:2014cfa}.%
\footnote{In contrast, $\int_{0}^{\infty} \!\! \df t \, \cI_{gj}(t,z,\ktsq,\mu)$ is ill-defined~\cite{Jain:2011iu,Gaunt:2014xxa}.}
As explained in \mycite{Gaunt:2014xxa} we can use \eq{I_int} to fix the $\delta(1-z)\, \delta(\ktsq)$ term in $\cI_{gg}(t,z,\ktsq,\mu)$ provided we know the result for $z<1$.
It equals the corresponding term in $\cI_{qq}(t,z,\ktsq,\mu)$ upon replacing $C_F \to C_A$ in the latter, see \mycite{Gaunt:2014cfa} for an explanation. There is no $\delta(1-z)$ (endpoint) term in $\cI_{gq}$.

As a consequence of \eq{ktsqlimits} the matching functions $\cI_{ij}(t,z,\ktsq, \mu)$ obey the same 
renormalization group equation (RGE) as the $\cI_{ij}(t,z,\mu)$~\cite{Jain:2011iu}:
\begin{align}
\label{eq:IRGE}
\mu\frac{\df}{\df\mu}\cI_{ij}(t,z,\ktsq, \mu)
= \sum_k \int\!\!\df t' \;  & \cI_{ik}(t-t', z,\ktsq, \mu) 
 \nn \\
& \convz  \Bigl[\gamma_B^i(t', \mu)\,  \delta_{kj} \delta(1-z) 
 - 2 \delta(t') P_{kj}(z,\mu) \Bigr]
 \,,
\end{align}
with the (Mellin) convolution $\,\convz\,$ defined in \eq{convz},
\begin{align}
\gamma_B^i(t, \mu)
&= -2 \Gamma^i_{\cusp}[\alpha_s(\mu)]\,\frac{1}{\mu^2}\cL_0\Bigl(\frac{t}{\mu^2}\Bigr) + \gamma_B^i[\alpha_s(\mu)]\,\delta(t)
\,,
\end{align}
and the collinear QCD splitting function $P_{ij}(z,\mu)$.
The latter are completely known to three loops~\cite{Moch:2004pa,Vogt:2004mw}. We give the LO expressions in \app{splittingfunc}.
The (virtuality-dependent) beam function anomalous dimension $\gamma_B^i(t, \mu)$ equals the jet function anomalous dimension, $\gamma_B^i(t, \mu) = \gamma_J^i(t, \mu)$~\cite{Stewart:2010qs}, which was derived for $i=g$ to three-loop order in \mycites{Becher:2009th,Berger:2010xi}.
Recently, the computation of the universal cusp anomalous dimension $\Gamma^i_{\cusp}$ to four loops in QCD has been completed~\cite{Moch:2018wjh,Grozin:2018vdn,Lee:2019zop,Henn:2019rmi,Bruser:2019auj,Henn:2019swt,vonManteuffel:2020vjv}. Thus all pieces in the dBF anomalous dimension necessary for N$^3$LL resummation are now available. 
The relevant cusp and non-cusp contributions to $\gamma_B^g(t, \mu)$ are collected up to three loops in the appendices of \mycites{Gaunt:2014xga,Gaunt:2014cfa} using our notation.

The first two terms ($n=0,1$)~\cite{Jain:2011iu} in the expansion of the dBF matching coefficient,
\begin{align}
\cI_{ij}(t,z,\ktsq,\mu) \,=\, \frac1\pi\, \sum_{n=0}^{\infty} \left(\dfrac{\alpha_s}{4\pi}\right)^{\!n} \cI_{ij}^{(n)}(t,z,\ktsq,\mu)
\,,\end{align}
are given in \app{oneloopIs}.
The structure of the two-loop coefficient is fixed by the RGE in \eq{IRGE} to be~\cite{Gaunt:2014xxa}
\begin{align} \label{eq:I2master}
\cI_{ij}^\two(t,z,\ktsq,\mu)&=
 \frac{1}{\mu^2} \cL_3\Bigl(\frac{t}{\mu^2}\Bigr) \frac{(\Gamma_0^i)^2}{2}\, \delta_{ij}\,\delta(1-z)\, \delta(\ktsq) \nn \\[1 ex] & 
  + \frac{1}{\mu^2} \cL_2\Bigl(\frac{t}{\mu^2}\Bigr)
  \Gamma_0^i \biggl[- \Bigl(\frac{3}{4} \gamma_{B\,0}^i + \frac{\beta_0}{2} \Bigr)\, \delta_{ij}\,\delta(1-z)\, \delta(\ktsq)  \nn\\&\qquad
  + 2P^\zero_{ij}(z)\, \delta(\ktsq) + P^\zero_{ij}(z)\, \delta\Bigl(t\frac{1-z}{z}-\ktsq\Bigr) \biggr]
  \nn \\[1 ex] &
  + \frac{1}{\mu^2} \cL_1\Bigl(\frac{t}{\mu^2}\Bigr)
  \biggl\{ \Bigl[\Gamma_1^i - (\Gamma_0^i)^2 \frac{\pi^2}{6} + \frac{(\gamma_{B\,0}^i)^2}{4} + \frac{\beta_0}{2} \gamma_{B\,0}^i \Bigr]\,\delta_{ij}\,\delta(1-z)\, \delta(\ktsq)
  \nn \\ & \qquad
  - \Big[\gamma_{B\,0}^i + 2 \Gamma_0^i \ln\!\Big(\frac{1-z}{z}\Big)\Big] P^\zero_{ij}(z)\,\delta(\ktsq) 
  + 2\Gamma_0^i\, I^\one_{ij}(z)\,\delta(\ktsq) \nn\\&\qquad
  - \Big[\gamma_{B\,0}^i + 2 \beta_0 + 2 \Gamma_0^i \ln\!\Big(\frac{1-z}{z}\Big)\Big] P^\zero_{ij}(z)\,\delta\Bigl(t\frac{1-z}{z}-\ktsq\Bigr)\nn\\&\qquad
  + 2\Gamma_0^i \Big[\theta \Big(t\dfrac{1-z}{z} - \ktsq\Big) \frac{1}{t} \cL_0\Bigl(\frac{\ktsq}{t}\Bigr)  + \frac{1}{t} \cL_0\Bigl(\frac{1-z}{z}-\frac{\ktsq}{t} \Bigr) \Big] P^\zero_{ij}(z) \nn \\[1 ex] & \qquad
  + 4 \sum_k  \Big[\delta\Bigl(t\frac{1-z}{z}-\ktsq\Bigr) P^\zero_{ik}(z) \Big] \conv_z P^\zero_{kj}(z) \biggr\} \nn \\[1 ex] & 
  + \frac{1}{\mu^2} \cL_0\Bigl(\frac{t}{\mu^2}\Bigr)\, 4 J^\two_{ij}(t,z,\ktsq)
  \,+\, \delta(t)\,\delta(\ktsq)\, 4 I^\two_{ij}(z)\,,
\end{align}
with the usual plus distributions
\begin{align} \label{eq:plusdefmaintxt}
\cL_n(x)
&= \biggl[ \frac{\theta(x) \ln^n x}{x}\biggr]_+
 = \lim_{\eps \to 0} \frac{\df}{\df x}\biggl[ \theta(x- \eps)\frac{\ln^{n+1} x}{n+1} \biggr]\,.
\end{align}
The anomalous dimension coefficients in the expression for $\cI_{gj}^\two$ are 
$\gamma_{B\,0}^g=2 \beta_0$, 
$\Gamma_0^g =4 C_A$, 
$\Gamma_1^g =4 C_A/3\big[(4-\pi^2)C_A+5\beta_0 \big]$, and
$\beta_0 = (11 C_A - 4 T_F n_f)/3$ with $n_f$ the number of (massless) quark flavours.
The coefficients of the $\delta(t) \delta(\ktsq)$ terms at one and two loops, $I^\one_{ij}(z)$ and $I^\two_{ij}(z)$, agree by virtue of \eq{I_int} with the $\delta(t)$ coefficients in the virtuality-dependent beam function and were computed in \mycites{Berger:2010xi,Gaunt:2014cfa}. The $I^\one_{gj}(z)$ can also be found in  \app{oneloopIs}.
We explicitly carry out the convolutions involving the one-loop splitting functions in the next-to-last line of \eq{I2master} for $i=g$ in \app{oneloopconvs}.

The novel results of this paper are the functions $J^\two_{gj}(t,z,\ktsq)$. We present the explicit expressions in the next section.
Although they multiply the $\mu$-dependent distribution $\cL_0(t/\mu^2)$ in \eq{I2master} they cannot be obtained by solving the RGE, because
\begin{align}
\mu \dfrac{d}{d\mu} \frac{1}{\mu^2} \cL_0\Bigl(\frac{t}{\mu^2}\Bigr)\, 4 J^\two_{ij}(t,z,\ktsq) 
&= -8 \delta(t) J^\two_{ij}(t,z,\ktsq) 
\nn\\ 
&= -8 \delta(t) \delta(\ktsq) \!\int_0^{\frac{1-z}{z}} \!\df r \, J^\two_{ij}(t,z, t\, r)\,.
\label{eq:Jderiv}
\end{align}
Hence, only the $\kt$ integral of the $J^\two_{ij}(t,z,\ktsq)$, but not the function itself, is fixed by the dBF anomalous dimension.
In \eq{Jderiv} we used the distributional identity~\cite{Gaunt:2014xxa}
\begin{align} \label{eq:distId}
\delta(t)\, \theta \Big(t\dfrac{1-z}{z} - \ktsq\Big) \, \dfrac{1}{t}\, f\bigg(\frac{\ktsq}{t}\bigg) 
= \delta(t)\,\delta(\ktsq) \int_0^{\frac{1-z}{z}} \!\!\! \df r f(r) \,,
\end{align}
which holds for any integrable function $f(\ktsq)$ as can be verified by integration over $\ktsq$.
The $\theta$-function in \eq{distId} implements the kinematic constraint in \eq{ktsqlimits} and is implicit in the $J^\two_{ij}(t,z,\ktsq)$ functions.

\section{Results for the dBF}
\label{sec:results}

Our calculation  of the two-loop dBF perfectly reproduces the known terms in \eq{I2master}, which serves as a strong cross check, and yields the following new results for the NNLO coefficient functions $J^\two_{gj}(t,z,\ktsq)$. 
Our notation is in close analogy to the one for the NNLO coefficients of the virtuality-dependent gluon beam function in \mycite{Gaunt:2014cfa}.
We define%
\footnote{Throughout this work $\theta(x) \delta(x) = \delta(x)$ is understood.}
\begin{align}
J_{gg}^\two(t,z,\ktsq) &=  \theta(z)\, \theta \Bigl(t\frac{1-z}{z} - \ktsq \Bigr)\Bigl[ C_A J_{ggA}^\two(t,z,\ktsq) + T_F n_f J_{ggF}^\two(t,z,\ktsq) \Bigr]
\,,  \label{eq:Jggdef}\\
J_{g q_i}^\two(t,z,\ktsq) &= J_{g \bar q_i}^\two(t,z,\ktsq) = \theta(z)\, \theta \Bigl(t\frac{1-z}{z} - \ktsq \Bigr)  C_F J_{gq}^\two(t,z,\ktsq)
\label{eq:Jgqdef}
\,,\end{align}
where $q_i$ ($\bar q_i$) denotes the (anti)quark of flavor $i$. 
Together with the finite support of the PDFs in \eq{dBi_OPE}  the two $\theta$-functions in \eqs{Jggdef}{Jgqdef} enforce the kinematic constraints $(1-x)/x \ge r\equiv \ktsq/t \ge 0$ and $1 \ge x \ge 0$. They also imply that taking the limits $z\to 1$ or $t\to0$ of the $J_{gi}^\two$ coefficients requires to integrate over the full range of $\ktsq$ (or equivalently $r$) first, see \mycite{Gaunt:2014xxa} for details.
The explicit expressions for $J_{ggA}^\two$, $J_{ggF}^\two$, and $J_{gq}^\two$ read%
\footnote{
All expressions for the $J^\two_{ij}$ as well as for the  $I^\two_{ij}$, which appear in \eq{I2master}, are available in electronic form upon request to the authors.
}
\begin{align}
J_{ggA}^\two &=
\beta_0\biggl\{
-\cL_0\Big(\frac{1-z}{z}-r\Big)\frac{p_{gg}(z)}{2t}
-\frac{1}{6} \delta \Bigl(t\frac{1-z}{z} - \ktsq \Bigr) \bigg[10 z^2 - 7 z - \frac{10}{(1-z)z} + 20 \bigg]
\nn\\
&\quad
- \delta (\ktsq) \bigg[
	\cL_1(1-z)
	-\frac{5}{3}\cL_0(1-z)
	+\bigg(\frac{14}{9}-\frac{\pi ^2}{6}\bigg) \delta (1-z)
	-\frac{\ln (1-z)}{1-z} 
	+ \frac{5}{3 (1-z)}
\bigg]
\nn\\
&\quad
+\frac{1}{2 t (1- z) (r+1)^2}\Big[
	4 r^4 (1-z) z^2
	-2 r^3 z \big(5 z^2-7 z+1\big)
	-r^2 \big(12 z^3-21 z^2+8 z-4\big) 
	\nn\\&\qquad
	-2 r \bigl(5 z^3-9 z^2+5 z-3\bigr)
	-4 z^3+7 z^2-5 z+3
\Big]
\biggr\}
\nn\\
&+C_A \Biggl\{
p_{gg}(z) \bigg[\frac{4}{t}\cL_1(r) + \frac{6}{t} \cL_1\Big(\frac{1-z}{z}-r\Big)
-  \frac{2}{t}  \cL_0\Big(\frac{1-z}{z}-r\Big) \ln \frac{1-z}{z}
 \bigg] 
  \nn\\
 &\quad 
 - \delta \Bigl(t\frac{1-z}{z} - \ktsq \Bigr) 
 \bigg[  p_{gg}(z) \biggl(2\ln^2 z + \frac{\pi^2}{6} \biggr)
 -\frac{8 z^4-25 z^3+33 z^2-16 z+ 8}{6 (1-z) z}  \bigg]
 \nn\\
 &\quad
 +\delta (\ktsq) \bigg[
	 6\cL_2(1-z)
	 +\bigg(\frac{4}{3}-2 \pi^2\bigg) \cL_0(1-z)
	  - \bigg(\frac{16}{9}-15 \zeta_3 \bigg) \delta (1-z)
	 +\frac{6 \pi ^2-4}{3 (1-z)}
	  	\nn\\&\qquad
	 -\frac{6\ln ^2(1-z)}{1-z}
 \bigg]
+\frac{2  \ln (r+1)}{(r+1)^4 t (1-z^2) z (1-r z-z)}
\biggl[
r^4 (1-z^2) z^2 \frac{p_{gg}(-z)}{2}
\nn\\&\qquad
+2 r^3 \bigl(2 z^6-2 z^5-2 z^3+2 z+1\bigr) 
+2 r^2 \bigl(3 z^6-9 z^5+3 z^4+z^2+6 z-1\bigr)
\nn\\&\qquad
+2 r \bigl(2 z^6-10 z^5+8 z^4-2 z^2+8	z-3\bigr)
+ z^6-7 z^5+9 z^4-3 z^3-z^2+9 z-6
\biggr]
 \nn\\
&\quad 
+\frac{2\ln (1-r z-z)}{r (r+1)^4 t (1-z) z} \Bigl[
	-2  z^3 r^7 (1-z) + 2 z^2  r^6 \big(5 z^2-6 z+1\big)
	\nn\\&\qquad
	+4 z r^5 \big(6 z^3-8 z^2+3 z-1\big) 
	+ r^4 \big(41 z^4-62 z^3+43 z^2-22 z+5\big)
	\nn\\&\qquad
	+r^3\big(53 z^4-88 z^3+87 z^2-48 z+15\big) 
	+3  r^2 \big(15 z^4-26 z^3+31 z^2-18z+7\big)
	\nn\\&\qquad
	+r\big(21 z^4-38 z^3+51 z^2-30 z+13\big)
	+ 2z(1-z) p_{gg}(z)
\Bigr]
\nn\\
&\quad
+\frac{2 \ln z}{(r+1)^2 t (1-z^2) (1-rz-z)} 
\Bigl[
	2 z^3 \bigl(z^2-1\bigr) r^5
	+4 z^2 r^4 \bigl(2 z^3-z^2-2 z+1\bigr) 
	\nn\\&\qquad	
	+2 z r^3\bigl(8 z^4-6 z^3-5 z^2+6 z-3\bigr) 
	+2 r^2 \bigl(12 z^5-11 z^4-2 z^3+9 z^2-8 z+3\bigr) 
	\nn\\&\qquad
	+r\bigl(22 z^5-26 z^4+5 z^3+18 z^2-19 z+10\bigr) 
	+8 z^5-12 z^4+5 z^3+7 z^2-9 z+5
\Bigr]
\nn\\
&\quad
+\frac{p_{gg}(z)}{r(r+1) t(1-r z-z)}\Bigl[
\bigl(2z r^2+6 z r-3 r +4 z-4\bigr)  \ln (1\!-\!z)
+ (3 r z+ 3z -2)  r  \ln r
\Bigr]
\nn\\
&\quad
+\frac{1}{2 (r+1)^4 t (1-z) z} \Bigl[
	12  z^3 r^6 (1-z)
	-6 z^2 r^5\bigl(11 z^2-13 z+3\bigr) 
	\nn\\&\qquad
	-z r^4 \bigl(148z^3-193 z^2+76 z-4\bigr) 
	-4 z r^3\bigl(43 z^3-59 z^2+31 z-3\bigr) 
	\nn\\&\qquad
	-2 r^2\bigl(54 z^4-73 z^3+55 z^2-13 z-2\bigr) 
	-2 r\bigl(17 z^4-19 z^3+19 z^2-16 z+8\bigr) 
	\nn\\&\qquad
	-4	z^4+z^3+6 z^2-10 z+4 \Bigr]
 	+p_{gg}(-z) \frac{\ln (1-r z)}{(r+1) t}
\Bigg\}\,,
\label{eq:JggA}
\end{align}

\begin{align}
J_{ggF}^\two &=
  \frac{2 C_F}{(r+1)^2t}\Bigg\{
P_{qg}(-r) P_{qg}\big[z(r+1)\big]
\ln \frac{1-rz- z}{z}
-16 r^4 z^2 -16 r^3  (3 z^2-z)
\nn\\
&\quad
-r^2 \bigl(52 z^2-32 z+2\bigr)-r \bigl(24 z^2-20 z+2\bigr)
-(2 z-1)^2
\Bigg\}\,,
\\[5 ex]
J_{gq}^\two &=
\beta_0 \,P_{gq}(z) \, \delta \Bigl(t\frac{1-z}{z} - \ktsq \Bigr)  \biggl( \ln z +\frac{5}{3}\biggr)
\nn\\
&
+C_F \Bigg\{P_{gq}(z)\bigg[
	-\delta \Bigl(t\frac{1-z}{z} - \ktsq \Bigr) \biggl( \ln ^2 z + 3\ln z +\frac{9}{2}
  	+ \frac{\pi^2}{3} \biggr)
  	+\frac{2}{t} \cL_1\Big(\frac{1-z}{z}-r\Big) 
\nn\\&\qquad  	
  	-\frac{3}{2t}\cL_0\Big(\frac{1-z}{z}-r\Big) 
  \bigg] 
 +\frac{2 z r^3+ r^2(4 z +2 )+r(z+6)-z+3}{(r+1)^2 t}  \ln \frac{z}{1-r z-z}
 \nn\\
 &\quad
 + \frac{8 z r^3+r^2(16 z +12)+r(5 z+18)-3 z+9}{2 (r+1)^2 t}
\Bigg\}
\nn\\
&+C_A \Bigg\{
P_{gq}(z) \bigg[\frac{4}{t}\cL_1(r)
+ \frac{4}{t}\cL_1\Big(\frac{1-z}{z}-r\Big) - \frac{2}{t} \cL_0\Big(\frac{1-z}{z}-r\Big) \ln \frac{1-z}{z}   
  \nn\\
&\quad
-\delta \Bigl(t\frac{1-z}{z} - \ktsq \Bigr)  \biggl(\ln^2z -\frac{7}{3}-\frac{\pi^2}{6} \biggr) \bigg]
+\frac{2 z}{t} \cL_0(r) 
+ P_{gq}(-z)  \frac{\ln (1-r z)}{(r+1) t}
  \nn\\
&\quad
+\frac{ \ln (1-r z-z)}{r (r+1)^4 t z} \Big[ 
	2 r^6 z^2+r^5 z (8 z-4)+r^4 \big(19 z^2-18 z+10\big) +r^3 \big(31 z^2-42 z+30\big)
	 \nn\\ &\qquad	
	+r^2 \big(31 z^2-50 z+42\big)
	+r \big(17 z^2-30 z+26\big) +4z P_{gq}(z)
	\Big]
  \nn\\
&\quad
+\frac{\ln(r+1)}{(r+1)^4 t z (1-r z-z)}  \Big[
	r^4 z^2 P_{gq}(-z) -2 r^3 \big(z^2-4 z-2\big)
	 \nn\\ &\qquad	
	+r^2 \big(6 z^3-6 z^2+24 z-4\big) + r \big(8 z^3-14 z^2+32 z-12\big)+3 z^3-8 z^2+18 z-12 
	\Big]
  \nn\\
&\quad
+\frac{2 \ln z}{(r+1)^2 t (1-r z-z)}
	\Big[
	r^4 z^2+3 r^3 (z-1) z+r^2 \big(5 z^2-8 z+4\big)+r \big(5 z^2-9 z+6\big)
	 \nn\\ &\qquad	
	+2 z^2-4 z+3
 	\Big]
+\frac{P_{gq}(z)}{r (r+1) t (1-r z-z)}\Big[ 
	\big(2 z r^2+r(6 z -3)+4 z-4\big) \ln (1-z) 
	 \nn\\ &\qquad	
	+\big(3 r z+3 z-2\big)r \ln r 
	\Big]
-\frac{1}{(r+1)^4 t z}\Big[
	4 r^5 z^2+2 r^4 z (8 z-1)+r^3 z (23 z-6)
	  \nn\\&\qquad
	+r^2 \big(13 z^2-15 z-2\big)
	+r \big(z^2-8 z+8\big)-z^2+3 z-2	
	\Big]
\Bigg\}\,.
\label{eq:Jgq}
\end{align}
Here we introduced the {\it unregulated} splitting function%
\footnote{Not to be confused with the proper (regulated) splitting function $P_{gg}(z)$ in \eq{Pij}.}
\begin{align}
p_{gg}(z) \equiv  \frac{2\bigl(z^2-z+1\bigr)^2}{(1-z)z}
\label{eq:pgg}
\end{align}
for compactness of notation.
The splitting function $P_{gq}(z)$ is given in \eq{Pij}.

We emphasize that \eqss{JggA}{Jgq} as well as \eq{I2master} are perfectly well-defined (in the distributional sense) as $z\to1$ or $t\to0$. 
Just like in \eq{distId} and its analog with $\delta(t)$ replaced by $\delta(1-z)$ one should keep in mind that taking these limits requires to integrate over $r\equiv \ktsq/t$ first. This is why these expressions in the present form contain terms which might seem to be unregularized (unintegrable) at first glance.

\section{Results for the generalized threshold beam function} 
\label{sec:genBF}

The beam function for generalized threshold factorization, $\tilde{B}_i(\tilde{t},x)$, is introduced and defined in \mycite{Lustermans:2019cau}. For $\tilde{t} \gg \Lambda^2_{\mathrm{QCD}}$ it can be written as the convolution of perturbative matching coefficients $\tilde{\mathcal{I}}_{ij}(\tilde{t},z)$ and the PDFs. The gluon matching coefficients  $\tilde{\mathcal{I}}_{gj}(\tilde{t},z)$ can be computed at two loops from the dBF results above via \eq{dBF2genBF}. The two-loop structure of $\tilde{\mathcal{I}}_{ij}(\tilde{t},z)$ is analogous to the one of $\mathcal{I}_{ij}(t,z)$~\cite{Gaunt:2014xga,Gaunt:2014cfa} and given in appendix~E
 of \mycite{Lustermans:2019cau}. It is completely fixed by the RGE and (known) lower-order ingredients, apart from the term proportional to $\delta(\tilde{t})$:%
\footnote{Note that our definition of $\tilde{I}^\two_{ij}(\tilde{t},z)$ is in analogy to $I^\two_{ij}(t,z)$ of \mycites{Gaunt:2014xga,Gaunt:2014cfa} and differs by a conventional factor of $4$ compared to the corresponding quantity in \mycite{Lustermans:2019cau}.}
\begin{align} \label{IstructuregenBF}
\tilde{\mathcal{I}}^\two_{ij}(\tilde{t},z) = \delta(\tilde{t})4\tilde{I}^\two_{ij}(\tilde{t},z) + \text{terms fixed by RGE and lower-order pieces}\,.
\end{align}

All terms in \eq{I2master}, when integrated according to \eq{dBF2genBF}, contribute to $\tilde{I}^\two_{ij}(\tilde{t},z)$, except those terms proportional to $\cL_n(t/\mu^2)\delta(\ktsq)$. Summing all contributions, one obtains
\begin{align}
\label{eq:tildeI2dtijmaster}
\tilde I^\two_{ij}(z)
={}&
\Biggl\{\Gamma^i_0 \biggl[ -\Li_3\Bigl(\frac{2 z}{1+z}\Bigr)+\biggl[\frac{\pi ^2}{6}-\frac12
\Li_2\Bigl(\frac{1-z}{1+z}\Bigr)\biggr] \ln \Bigl(\frac{2 z}{1+z}\Bigr)-\frac{1}{2} \ln \Bigl(\frac{1-z}{1+z}\Bigr) \ln^2\Bigl(\frac{2	z}{1+z}\Bigr)
\nn\\
&
+\zeta_3 \biggr] - \Big(\frac{\beta_0}{4}+ \frac{\gamma^i_{B0}}{8}\Big) \ln^2\Bigl(\frac{2 z}{1+z}\Bigr)
\Biggr\}
P^\zero_{ij}(z) \
+ I^\two_{ij}(z)
+\Delta\tilde I^\two_{ij}(z)
\,.
\end{align}
The term $\Delta\tilde I^\two_{ij}(z)$ corresponds to the contribution from integrating the terms 
$\propto \cL_0(t/\mu^2)J_{ij}$ and $\propto \cL_1(t/\mu^2) P_{ik}\otimes P_{kj}$ in \eq{I2master}. We computed the $\Delta\tilde I^\two_{gj}(z)$ using our results from \sec{results} and \app{oneloopconvs}. The $\Delta\tilde I^\two_{gj}(z)$ are regular functions in $z$, i.e.\ they do not involve distributions.  
We give the explicit expressions, expressed in terms of ordinary polylogarithms up to weight three,  in \app{genBFapp}.
We emphasize that the linearly polarized term $\propto  \mathscr{B}_g$ in \eq{dBg_tensor} vanishes upon the $\kt$ integration in \eq{dBF2genBF}. Our two-loop results for the gluon generalized threshold beam function are therefore complete.

\section{Conclusions} 
\label{sec:conclusions}

The fully-differential beam function $B_j(t,x,\ktsq,\mu)$ has many potential applications in high-energy collider physics, from resummations of multi-differential observables, to IR subtraction schemes and improved parton showers. 
In this paper we utilised the methods we developed in  \mycites{Gaunt:2014xga, Gaunt:2014cfa, Gaunt:2014xxa} to compute the two-loop matching coefficients $\cI_{gj}^\two(t,z,\ktsq,\mu)$ between the unpolarised gluon dBF $B_g(t,x,\ktsq,\mu)$ and the usual PDFs $f_j(x,\mu)$.
Our results are an important ingredient to obtain the full NNLO singular contributions as well as the NNLL$'$ and N$^3$LL resummation for observables that probe both the virtuality and the transverse momentum of the colliding gluons. From our two-loop dBF results, we also obtained, for the gluon case, the complete two-loop matching coefficients between the beam function appearing in generalized threshold factorisation \cite{Lustermans:2019cau} and the PDFs.

\begin{acknowledgments}
JRG acknowledges the University of Freiburg for hospitality, where a part of this work was completed.
\end{acknowledgments}

\appendix

\section{Tree-level and one-loop matching coefficients}
\label{app:oneloopIs}

We expand the beam function matching coefficient as
\begin{align}
\cI_{ij}(t,z,\ktsq,\mu) \,=\, \frac1\pi\, \sum_{n=0}^{\infty} \left(\dfrac{\alpha_s}{4\pi}\right)^{\!n} \cI_{ij}^{(n)}(t,z,\ktsq,\mu)
\,.\end{align}
At tree level we have
\begin{equation}
\label{eq:Iijtree}
\cI_{ij}^\zero(t,z,\ktsq,\mu) = \delta_{ij}\,\delta(t)\,\delta(\ktsq)\,\delta(1-z)
\,.\end{equation}
The one-loop matching coefficients are~\cite{Jain:2011iu}
\begin{align} \label{eq:Iijoneloop}
\cI_{ij}^\one(t,z,\ktsq,\mu)
&= \frac{1}{\mu^2} \cL_1\Bigl(\frac{t}{\mu^2}\Bigr) \Gamma_0^i\, \delta_{ij}\delta(\ktsq)\delta(1-z)
  + \frac{1}{\mu^2} \cL_0\Bigl(\frac{t}{\mu^2}\Bigr)
  \Bigl[- \frac{\gamma_{B\,0}^i}{2}\,\delta_{ij}\delta(\ktsq)\delta(1-z) \nn \\ & \quad
+ 2\, \delta\Big(t \frac{1-z}{z}-\ktsq\Big) P_{ij}^\zero(z) \Bigr]
  + \delta(t)\,\delta(\ktsq)\, 2I_{ij}^\one(z)
\,,\end{align}
%
with the $\mu$-independent one-loop constants
\begin{align}
I_{q_iq_j}^\one(z) &= \delta_{ij}\, C_F\, \theta(z) I_{qq}(z)
\,,\nn\\
I_{q_ig}^\one(z) &= T_F\, \theta(z) I_{qg}(z)
\,,\nn\\
I_{gg}^\one(z) &= C_A\, \theta(z) I_{gg}(z)
\,,\nn\\
I_{gq_i}^\one(z) &= C_F\, \theta(z) I_{gq}(z)
\,.\end{align}
The one-loop gluon matching functions~\cite{Berger:2010xi} are given by\footnote{Here $I_{ij}(z) \equiv \cI_{ij}^{(1,\delta)}(z)$ in the notation of \mycites{Stewart:2010qs, Berger:2010xi}.}
\begin{align} \label{eq:Igdel_results}
I_{gg}(z)
&= \cL_1(1-z)\,\frac{2(1-z + z^2)^2}{z} - \frac{\pi^2}{6} \delta(1-z) - P_{gg}(z) \ln z
\,, \nn \\
I_{gq}(z)
&= P_{gq}(z)\ln \frac{1-z}{z} + \theta(1-z) z
\,.\end{align}

\section{Perturbative ingredients}
\label{app:pert}

\subsection{Splitting functions} 
\label{app:splittingfunc}

We expand the PDF anomalous dimensions ($\gamma^f_{ij}=2 P_{ij}$) in the $\overline{\mathrm{MS}}$ scheme as
\begin{align}
\label{eq:Pijexp}
P_{ij}(z,\alpha_s) = \sum_{n=0}^{\infty} \left(\dfrac{\alpha_s}{2\pi}\right)^{n+1} P_{ij}^{(n)}(z)\,.
\end{align}
The one-loop terms read
\begin{align}
P_{q_i q_j}^\zero(z) &= C_F\, \theta(z)\, \delta_{ij} P_{qq}(z)
\,,\nn\\
P_{q_ig}^\zero(z) = P_{\bar q_ig}^\zero(z) &= T_F\, \theta(z) P_{qg}(z)
\,,\nn\\
P_{gg}^\zero(z) &= C_A\, \theta(z) P_{gg}(z) + \frac{\beta_0}{2}\,\delta(1-z)
\,,\nn\\
P_{gq_i}^\zero(z) = P_{g\bar q_i}^\zero(z) &= C_F\, \theta(z) P_{gq}(z)
\,,\end{align}
with the usual one-loop (LO) quark and gluon splitting functions
\begin{align} \label{eq:Pij}
P_{qq}(z)
&= \cL_0(1-z)(1+z^2) + \frac{3}{2}\,\delta(1-z)
\equiv \biggl[\theta(1-z)\,\frac{1+z^2}{1-z}\biggr]_+
\,,\nn\\
P_{qg}(z) &= \theta(1-z)\bigl[(1-z)^2+ z^2\bigr]
\,,\nn\\
P_{gg}(z)
&= 2 \cL_0(1-z) \frac{(1 - z + z^2)^2}{z}
\,,\nn\\
P_{gq}(z) &= \theta(1-z)\, \frac{1+(1-z)^2}{z}
\,.\end{align}

\subsection{Convolutions of one-loop functions} \label{app:oneloopconvs}

The (Mellin) convolution of two functions (of light-cone momentum fractions) is defined as
\begin{align}
\label{eq:convz}
f(z) \convz g(z) = \int_z^1\! \frac{\df w}{w}\, f(w) g \Bigl(\frac{z}{w}\Bigr)
\,.\end{align}
%
For $i=g$ and $j=g,q$ the following non-trivial convolutions of one-loop splitting functions appear  in the two-loop dBF matching coefficient, \eq{I2master} (with $r \equiv \ktsq/t \ge 0$, $1\ge z \ge 0$):
\begin{align}
\label{eq:convPggPgg}
&\Big[\delta\Bigl(t\frac{1-z}{z}-\ktsq\Bigr) P_{gg}(z) \Big]  \conv_z   P_{gg}(z) = 
\delta (\ktsq) \biggl[ 8 \cL_1(1-z)-\frac{2\pi ^2 }{3} \delta (1-z)-\frac{8 \ln (1-z)}{1-z} \biggr]
\nn\\
&\hspace{5 ex}
+2p_{gg}(z) \biggl[ \theta\Bigl(\frac{1-z}{z}-r\Bigr) \frac{1}{t} \cL_0(r)+ \frac{1}{t} \cL_0\Bigl( \frac{1-z}{z}-r \Bigr) + \delta\Bigl(t\frac{1 - z}{z}-\ktsq\Bigr)  \ln (z) \biggr]
\nn\\
&\hspace{5 ex}
+ \frac{4}{(r+1)^4 t z}\Bigl[
-r^6 z^3 - r^5 \big(5 z^3-z^2\big)-2 r^4 z \big(6 z^2-2 z+1\big)-2 r^3 z \big(9 z^2-4 z+3\big)
\nn\\&\hspace{5 ex}\quad
-r^2 \big(17 z^3-8 z^2+7 z+2\big)-3 r \big(3 z^3-z^2+z+1\big)-2 \big(z^3+1\big)
	\Bigr] \theta\Bigl(\frac{1-z}{z}-r\Bigr)
\,, \\
&\Big[\delta\Bigl(t\frac{1-z}{z}-\ktsq\Bigr) P_{gq}(z) \Big]  \conv_z   P_{qg}(z) =  
\frac{P_{qg}(-r) P_{qg}\big[z(r+1)\big] }{(r+1)^2 t} 
\,, \\[1 ex]
&\Big[\delta\Bigl(t\frac{1-z}{z}-\ktsq\Bigr) P_{gg}(z) \Big]  \conv_z   P_{gq}(z) = 
\nn\\
&\hspace{22 ex}\;
\frac{2}{t} P_{gq}(z)  \biggl[ \cL_0(r) -\frac{1}{r}\biggr] \theta\Bigl(\frac{1-z}{z}-r\Bigr)
- \frac{p_{gg}(-r)}{(r+1)^2 t} P_{gq}\big[z(r+1)\big]
\label{eq:convPggPgq}
\,, \\[1 ex]
&\Big[\delta\Bigl(t\frac{1-z}{z}-\ktsq\Bigr) P_{gq}(z) \Big]  \conv_z   P_{qq}(z)  =  
\nn\\
&\quad
 P_{gq}(z) \biggl(2 \ln z+\frac{3}{2}\biggr) \,\delta \Bigl(t \frac{1-z}{z} -\ktsq\Bigr)
+\frac{P_{qg}(-r)}{(r+1)^2 tz}   \bigl[(r+1)^2 z^2+1\bigr] \cL_0\Bigl( \frac{1-z}{z}-r \Bigr)
\,.
\label{eq:convPgqPqq}
\end{align}
Here we employed the definitions in \eqs{pgg}{Pij}.
Note that the two terms in \eq{convPggPgq} combine to a expression regular in the limit $r\to0$.

\section{Expressions for the $\Delta\tilde I^\two_{gj}$}
\label{app:genBFapp}
In analogy to \eqs{Jggdef}{Jgqdef} we define
\begin{align}
\Delta\tilde I^\two_{gg}(z) &=  \theta(z)\, \theta (1-z) \Bigl[ C_A \Delta\tilde I^\two_{ggA}(z) + T_F n_f \Delta\tilde I^\two_{ggF}(z) \Bigr]
\,,  \\
\Delta\tilde I^\two_{gq_i}(z) &= \Delta\tilde I^\two_{g\bar q_i}(z) = \theta(z)\, \theta (1-z)\,  C_F \Delta\tilde I^\two_{gq}(z)
\,.
\end{align}
We find%
\footnote{
	All expressions for the $\Delta\tilde I^\two_{gj}$ are available in electronic form upon request to the authors.
}
\begin{align}
& \Delta\tilde I^\two_{ggA}(z) =\beta_0 \Biggl\{
	 \frac{(21 z+16) \ln \left(\frac{2 z}{z+1}\right) z^2}{6 (z-1)}
	 +\frac{1}{2} (z+1) \ln ^2(z)
	 +\biggl[ \frac{-39 z^2+21 z-23}{6 (z-1)z} 
\nn\\&\quad
	 	+ \ln (2) (z+1)\biggr] \ln (z)
	+p_{gg}(z) \biggl[
	\frac{1}{4} \ln ^2\left(\frac{2 z}{z+1}\right)
	+ \frac12 \ln\left( \frac{z}{1-z}\right) \ln \left(\frac{2z}{z+1}\right)
\nn\\&\quad	
		+\frac{1}{2} \ln (z) \ln \left(\frac{1-z}{z+1}\right)-\frac{\pi ^2}{8}
	\biggr] 
	+\left(z-\frac{p_{gg}(z)}{2}+1\right)\Li_2(-z)
	+\frac{1}{2} p_{gg}(z) \Li_2(z)
\nn\\&	
	-\frac{1}{2} p_{gg}(z) \Li_2\left(\frac{z-1}{z+1}\right)
	+\frac{-155 z^3+3 \pi ^2 z^2+114 z^2+3 \pi ^2	z-48 z+71}{36 z}
\nn\\&	
	+\frac{42 z^2-24 z+23}{6 (z-1) z} \ln \left(\frac{z+1}{2} \right)
\Biggr\}
\nn\\[2 ex]&	
+C_A \Bigg\{
	\frac{2 \left(7 z^4+8 z^3-7 z^2-8 z+1\right)}{3 z (z^2-1) }  \ln ^3(z)
	+\biggl[
		\frac{-27 z^3+18 z^2-30 z+11}{3 z}
	\nn\\&\quad
		-\frac{2 \left(z^4-2 z^3-z^2+2 z+1\right) \ln(2)}{z^2-1}
	\biggr] \ln ^2(z)
	+\biggl[
		-4 \ln ^2(2) (z+1)
		-\frac{4}{3} \left(11 z^2+12\right) \ln (2)
	\nn\\&\quad	
		+\frac{1}{9 z (z^2-1)} \Big( 
			9 \pi ^2 z^5-91 z^5-24 \pi ^2 z^4-111 z^4-15 \pi ^2 z^3+7	z^3+24 \pi ^2 z^2+80 z^2
		\nn\\&\qquad	
			+15 \pi ^2 z+84 z+6 \pi ^2+31 \Big)
	\biggr] \ln (z)
	+\frac{5 z^4-14 z^3+15 z^2-6 z+5 }{3 (z-1) z} \ln^3\left(\frac{2z}{z+1}\right)
\nn\\&	
	+\frac{31 z^5-31 z^4+27 z^3+31 z^2-27 z+27 }{3 z \left(z^2-1\right)} \ln^3(z+1)
\nn\\&		
	-\frac{3 \left(z^4-2 z^3+3 z^2-2 z+1\right)	}{(z-1) z}	\ln \left(\frac{2 z}{1-z}\right)
	\ln ^2\left(\frac{2 z}{z+1}\right)
\nn\\&
	+\ln^2(z+1) \biggl[-\frac{2 (z-1) (z+1)^2}{z}
		+\frac{-11 z^5+9 z^4-7 z^3-9 z^2+7 z-5}{z \left(z^2-1\right)} 	\ln (z)
	\nn\\&\quad	
		-\frac{2 \left(11 z^5-11 z^4+9 z^3+11 z^2-9 z+9\right) \ln (2)}{z \left(z^2-1\right)}
	\biggr] 
\nn\\&
	+\biggl[\frac{-z^4-2z^3-3 z^2+6 z-1}{(z-1) z}  \ln^2(z)
	+\frac{2 \pi ^2}{3}\biggr] \ln \left(\frac{2 z}{1-z}\right)
\nn\\&
	+\biggl[
		\frac{-z^4-2 z^3-3 z^2+6 z-1}{(z-1) z} \ln^2(z)
		+\frac{2 \left(z^4+2 z^3+3 z^2-6 z+1\right) }{(z-1) z} \ln \left(\frac{2 z}{1-z}\right) \ln (z)
	\nn\\&\quad	
		+\frac{-6 \pi ^2 z^4+217 z^4+60 \pi ^2 z^3+167 z^3-18
		\pi ^2 z^2-180 z^2-36 \pi ^2 z-71 z-6 \pi ^2-211}{18 (z-1) z}
	\biggr]
	\nn\\&\quad		
	 \times \ln \left(\frac{2 z}{z+1}\right)
	+\biggl[
		\frac{4 (z-1) \ln (2) (z+1)^2}{z}
		+\frac{z^5+5z^4+z^3-5 z^2-z-5 }{z \left(z^2-1\right)} \ln ^2(z)
	\nn\\&\quad	
		-\frac{\pi ^2 \left(5 z^5-10 z^4-3 z^3+10 z^2+3 z+2\right)}{3 z \left(z^2-1\right)}
		+\biggl(
			\frac{4 \left(3 z^5-z^4+3 z^3+z^2-3 z+1\right) \ln (2)}{z \left(z^2-1\right)}
		\nn\\&\qquad	
			+\frac{2 (z+1) \left(5 z^2+z+23\right)}{3 z}
		\biggr) \ln (z)
	+\frac{\left(11 z^4-26 z^3+33 z^2-18z+11\right) \ln ^2(2)}{(z-1) z}
	\biggr] \ln (z+1)
\nn\\&
	+\ln (1-z) \biggl[
		-\frac{4 (z-1) \ln (2) (z+1)^2}{z}+\frac{91 z^3-6 \pi ^2 z^2-45 z^2+45 z-91}{9 z}
	\nn\\&\quad		
		+\bigg(\frac{4(z-1) (z+1)^2}{z}+\frac{8 \left(z^4-3 z^3+3 z^2-z+1\right)}{(z-1) z} \ln (z) \bigg) \ln (z+1)
	\biggr]
\nn\\&
	+\ln (z+1) p_{gg}(-z) \Li_2(-2 z)
	+\biggl[
		-\frac{2\left(17 z^3-6 z^2-23\right)}{3 z}
		-8 (z+1) \ln (1-z)
	\nn\\&\quad	
		-\frac{2 \left(5 z^5-13 z^4+z^3+13 z^2-z+9\right) }{z \left(z^2-1\right)} \ln (z)
		+\biggl(-2 \ln \left(\frac{2z}{z+1}\right)-7 \ln (z+1)\biggr) p_{gg}(z)
	\nn\\&\quad	
		-\frac{4 \left(3 z^5-4 z^4+3 z^3+4 z^2-3 z+4\right) \ln (2)}{z \left(z^2-1\right)}
	\biggr]\Li_2(-z)
	+\biggl[
		\frac{2 \left(6 z^3+18 z^2+3 z+11\right)}{3 z}
	\nn\\&\quad	
		+\left(-\ln (z)-5 \ln \left(\frac{2 z}{z+1}\right)\right) p_{gg}(z)
	\biggl]\Li_2(z)
	+\biggl[ 
		\Big( 4 \ln (z)-3 \ln (z+1)+3 \ln (2) \Big) p_{gg}(z)
	\nn\\&\quad	
		-\frac{4 (z-1) (z+1)^2}{z} 
	\biggr] \Li_2\left(\frac{z-1}{z+1}\right)
	+\biggl[2 \ln \left(\frac{2 z}{z+1}\right)-\ln (2)\biggr] p_{gg}(-z) \Li_2\left(-\frac{z}{z+1}\right)
\nn\\&
	+\ln (z+1) p_{gg}(-z) \Li_2\left(\frac{1}{2	z+2}\right)
	+\frac{2 \left(z^4+2 z^3+3 z^2-6 z+1\right)}{(z-1) z} \Li_3\left(\frac{1-z}{2}\right) 
\nn\\&
 	-\frac{2 \left(z^4+2 z^3+3 z^2-6 z+1\right)}{(z-1) z} \Li_3(1-z)
	+\biggl[\frac{4 \left(z^4-4 z^3+3 z^2+1\right)}{(z-1) z}
	+2 p_{gg}(z)\biggr] \Li_3\left(\frac{z-1}{2 z}\right)
\nn\\&	
	-p_{gg}(-z)\Li_3(-2 z) + 3 p_{gg}(-z) \Li_3(-z) 
	+ 3 p_{gg}(z) \Li_3(z)
	+p_{gg}(-z) \Li_3\left(\frac{1}{z+1}\right)
\nn\\&		
	-3 p_{gg}(-z)\Li_3\left(-\frac{z}{z+1}\right) 
	+ 11 p_{gg}(z) \Li_3\left(\frac{2 z}{z+1}\right)
	-3 p_{gg}(z) \Li_3\left(\frac{z+1}{2}\right)
\nn\\&			
	+p_{gg}(-z)\Li_3\left(\frac{1}{2 z+2}\right)
	+p_{gg}(z) \biggl[
		\frac{2}{3} \ln ^3\left(\frac{2 z}{z+1}\right)
		+\left(\frac{9}{2} \ln (1-z)-\ln \left(\frac{2z}{1-z}\right)-\ln (2)\right) 
	\nn\\&\quad	
		\times \ln ^2\left(\frac{2 z}{z+1}\right)
		-\ln \left(\frac{1-z}{z+1}\right) \ln ^2\left(\frac{2 z}{z+1}\right)
		+ \ln\left(\frac{2 z}{z+1}\right) \biggl(
			-\ln ^2(z)-9 \ln(1-z) \ln (z)
		\nn\\&\qquad	
			-3 \ln (z+1) \ln (z)+\ln (2) \ln (z)+\ln ^2\left(\frac{2 z}{1-z}\right)-2 \ln (2) \ln \left(\frac{2 z}{1-z}\right)+\ln ^2(2)-\pi ^2
		\biggr)
	\nn\\&\quad	
		+\ln (1-z) \left(\frac{11}{2} \ln^2(z) -2 \ln (z+1) \ln (z)+6 \ln (2) \ln (z)\right) -2 \zeta_3
	\biggr]
\nn\\&	
	+\frac{1}{108 z \left(z^2-1\right)}\bigg[
		2403 \zeta_3 z^5-174\pi ^2 z^5-907 z^5-1485 \zeta_3 z^4-180 \pi ^2 z^4+360 z^4+2403 \zeta_3 z^3
	\nn\\&\quad			
		+138 \pi ^2 z^3+937 z^3+1485 \zeta_3 z^2+186 \pi ^2 z^2+319 z^2-2403 \zeta_3 z
		+36	\pi ^2 z-30 z+1485 \zeta_3
	\nn\\&\quad					
		-6 \pi ^2-679 
	\bigg]
	+\frac{\left(-z^5+7 z^4+3 z^3-7 z^2-3 z-3\right) \ln ^3(2)}{3 z \left(z^2-1\right)}
	-\frac{2 (z-1)(z+1)^2 \ln ^2(2)}{z}
\nn\\&	
	+\frac{\left(4 \pi ^2 z^5-11 \pi ^2 z^4-6 \pi ^2 z^3+11 \pi ^2 z^2+6 \pi ^2 z+3 \pi ^2\right) \ln (2)}{3 z \left(z^2-1\right)}
\Bigg\}
\end{align}


\begin{align}
&\Delta\tilde I^\two_{ggF}(z) =
\frac{C_F}{z} \Biggl\{
\frac{2}{3} (z+1) \left(16 z^2+5 z+4\right) \Li_2\left(\frac{z-1}{z+1}\right)
-\frac{2}{3} \left(16 z^3+21 z^2+12 z+4\right)\Li_2(z)
\nn\\&
	+\Li_2(-z) \biggl[\frac{4}{3} \left(8 z^3+6 z^2+3 z+2\right)
	-4 z (z+1) \ln \left(\frac{z}{1-z}\right)\biggr]
	-4 z (z+1)\Li_3\left(\frac{1-z}{2}\right)
\nn\\&	
	+4 z (z+1) \Li_3(1-z)
	+4 z (z+1) \Li_3\left(\frac{z-1}{2 z}\right)
	+\frac{1}{3} (z+1) \left(16 z^2+5 z+4\right) \ln^2(z+1)
\nn\\&	
	+\frac{1}{3} (z+1) \left(16 z^2+5 z+4\right) \ln ^2(2)
	+\ln \left(\frac{z}{1-z}\right) \biggl[ \frac{2}{3} (z+1) \left(16 z^2+5 z+4\right) \ln (z+1)
\nn\\&\quad	
	-\frac{1}{3}\pi ^2 z (z+1) \biggr]
	+6 z \ln (2) \ln \left(\frac{1-z}{z+1}\right)
	+\frac{1}{27} \Bigl(72 \pi ^2 z^3+725 z^3+81 \pi ^2 z^2 +57 z^2+45 \pi ^2 z
\nn\\&\quad		
	-741 z+18 \pi^2-41 \Bigr)
	-\biggl[\frac{2}{3} \left(8 z^3+15 z^2+9 z+2\right) + 4 z (z+1) \ln (2)\biggr] \ln ^2(z)
\nn\\&		
	+\ln (1-z) \biggl[-\frac{2}{9} \left(37 z^3-3 z^2-21z-13\right)
	+\frac{2}{3} \left(16 z^3+21 z^2+4\right) \ln (2)+2 z (z+1) \ln ^2(z)
\nn\\&\quad		
	+4 z (z+1) \ln (2) \ln (z)\biggr]
	+\biggl[-\frac{2}{9} \left(83 z^3+174 z^2+75z+7\right)-2 z (3 z+1) \ln (2)\biggr] \ln (z)
\nn\\&		
	+\biggl[\frac{2}{3} \left(40 z^3+57 z^2+15 z-2\right)-\frac{2}{3} \left(16 z^3+21 z^2+4\right) \ln (2)\biggr] \ln(z+1)
	-2 z (z+1) \ln ^3(z)
\nn\\&			
	-\frac{2}{3} (z+1) \left(40 z^2+17 z-2\right) \ln (2)
	 \Biggr\}
\end{align}


\begin{align}
&\Delta\tilde I^\two_{gq}(z) =
\beta_0 \left(\ln (z)+\frac{5}{3}\right) \ln \left(\frac{2 z}{z+1}\right)P_{gq}(z)
\nn\\&	
+\frac{C_A}{z} \Biggl\{
	\frac{2}{3} \left(z^2+9 z-2\right) \ln ^3(z)
	+\biggl[2 \ln (2) (2 z-1)+\frac{1}{6} \left(-16 z^3-12 z^2-72 z+9\right)\biggr] \ln ^2(z)
\nn\\&
	+\frac{1}{2}	\left(-z^2-10 z+2\right) \ln \left(\frac{2 z}{1-z}\right) \ln ^2(z)
	+ \ln (z) \biggl[
		-2 \ln ^2(2) \left(2 z^2+z+1\right)
	\nn\\&\quad
		+\frac{1}{18} \left(88 z^3-15 \pi ^2 z^2+15 z^2-18 \pi^2 z+186 z-12 \pi ^2+45\right)
		-\frac{1}{3} z \left(8 z^2+15 z+48\right) \ln (2)
	\biggr]
\nn\\&
	+\frac{1}{6} \left(-7 z^2+2 z-10\right) \ln ^3\left(\frac{2	z}{z+1}\right)
	+\frac{1}{6} \left(-27 z^2+50 z-50\right) \ln ^3(z+1)
	+ \ln^2(z+1) \biggl[
		\ln (2) z^2
	\nn\\&\quad
		+\frac{1}{6} \left(4 z^3-3 z^2+24 z+25\right)
		+\frac{1}{2} \left(5 z^2-10z+6\right) \ln (z)
	\biggr]
	+\ln \left(\frac{2 z}{z+1}\right) \biggl[
		2 \ln ^2(2) z^2
	\nn\\&\quad
		+\frac{1}{2} \left(-z^2-10 z+2\right) \ln ^2(z)
		+\frac{1}{6} \pi ^2 \left(5 z^2+14	z+2\right)
	+\left(z^2+10 z-2\right) \ln (z) \ln \left(\frac{2 z}{1-z}\right)
	\biggr] 
\nn\\&
	+\ln (z+1) \biggl[
		\frac{1}{2} \left(7 z^2-2 z+14\right)\ln ^2(z)
		+\Bigl(8 \ln (2) z	+\frac{1}{3} \left(8 z^3+9 z^2+60 z+59\right)\Bigr) \ln (z)
	\nn\\&\quad	
		+\frac{1}{9} \left(-44 z^3+9 \pi ^2 z^2-108 z^2+9 \pi ^2 z-153 z+6 \pi^2-89\right)
		+\frac{1}{2} \left(-7 z^2+10 z-18\right) \ln ^2(2)
	\nn\\&\quad	
		+\frac{1}{3} \left(-4 z^3+3 z^2-24 z-25\right) \ln (2)
	\biggr] 
	+ \biggl[
		2 \ln (z)z^2
		+\frac{1}{6} \left(-\pi ^2 z^2+23 z^2-4 \pi ^2 z+14 z-37\right)
	\nn\\&\quad	
		+\left(\frac{1}{3} \left(-4 z^3-3 z^2-24 z-25\right)
		-2 \left(3 z^2+4\right) \ln (z)\right) \ln(z+1)
	\nn\\&\quad		
		+\frac{1}{3} \left(4 z^3+3 z^2+24 z+25\right) \ln (2)
	\biggr] \ln (1-z)
	+z \ln (z+1)P_{gq}(-z) \Li_2(-2 z)
\nn\\&
	+\Li_2(-z) \biggl[
		\frac{1}{3} \left(-6 z^2+12 z+59\right)
		-2 z (z+4) \ln (1-z)
		+\left(11 z^2-2 z+18\right) \ln (z)
	\nn\\&\quad	
		+\Bigl(-2 \ln (1-z) z-5 \ln (z+1) z-2 \ln (2) z \Bigr)P_{gq}(z)
		+4 \left(2 z^2-3 z+4\right) \ln (2)
	\biggr] 
\nn\\&
	+ \biggl[
		\frac{1}{3} \left(-4 z^3+9 z^2-6 z+9\right)+\Bigl(2 \ln (1-z) z-8 \ln (z) z+5 \ln (z+1) z
		-5 \ln (2) z \Bigr)P_{gq}(z)
	\biggr] 
	\nn\\&\quad	
	\times \Li_2(z)
	+\Li_2\left(\frac{z-1}{z+1}\right) \biggl[
		\Bigl(-2 \ln (1-z) z+6 \ln (z) z-3 \ln (z+1) z+3 \ln (2) z \Bigr) P_{gq}(z)
	\nn\\&\quad		
		+\frac{1}{3} \left(4 z^3-3 z^2+24 z+25\right)
	\biggr] 
	+\biggl[2 z \ln \left(\frac{2 z}{z+1}\right)-z \ln (2)\biggr]P_{gq}(-z) \Li_2\left(-\frac{z}{z+1}\right)
\nn\\&
	+z \ln(z+1)P_{gq}(-z) \Li_2\left(\frac{1}{2 z+2}\right)
	+\left(z^2+10 z-2\right) \biggl[ \Li_3\left(\frac{1-z}{2}\right)
	- \Li_3(1-z) \biggr]
\nn\\&	
	-4	\left(z^2+z+1\right) \Li_3\left(\frac{z-1}{2 z}\right)
	-zP_{gq}(-z) \Li_3(-2 z)+3 zP_{gq}(-z) \Li_3(-z)
	+3 z P_{gq}(z) \Li_3(z)
\nn\\&		
	+z	P_{gq}(-z) \Li_3\left(\frac{1}{z+1}\right)
	-3 zP_{gq}(-z) \Li_3\left(-\frac{z}{z+1}\right)
	+11 zP_{gq}(z) \Li_3\left(\frac{2z}{z+1}\right)
\nn\\&		
	-3 zP_{gq}(z) \Li_3\left(\frac{z+1}{2}\right)
	+zP_{gq}(-z) \Li_3\left(\frac{1}{2 z+2}\right)
	+P_{gq}(z) \biggl[
		\frac{1}{3} z \ln^3(z)
		+z \ln (2) \ln ^2(z)
	\nn\\&\quad	
		+2 z \ln ^2(1-z) \ln (z)
		-\frac{1}{6} \left(-14+\pi ^2\right) z \ln (z)
		-7 z \ln (1-z) \ln \left(\frac{2 z}{z+1}\right) \ln(z)
	\nn\\&\quad	
		-\frac{1}{3} z \ln ^3(z+1)
		+\left(\frac{7}{2} z \ln (1-z)+\frac{3}{2} z \ln \left(\frac{2 z}{1-z}\right)\right) \ln ^2\left(\frac{2 z}{z+1}\right)
		+ \ln ^2(z+1) \Bigl(4 \ln (z) z 
	\nn\\&\qquad					
			+8 \ln (2) z \Bigr) 
		+\left(-2 z \ln ^2(z)-4 z \ln (2) \ln (z)
		+\frac{1}{3} \left(-7+2 \pi ^2\right) z+z \ln ^2(2)\right) \ln (z+1)
	\nn\\&\quad		
		+\ln (1-z)	\left(\frac{3}{2} z \ln ^2(z)-2 z \ln (z+1) \ln (z)+4 z \ln (2) \ln (z)-\frac{\pi ^2 z}{2}\right)
		-4 z \zeta_3
		-\frac{2}{3} z \ln ^3(2)
	\nn\\&\quad				
		-\frac{1}{3} \left(-7+2	\pi ^2\right) z \ln (2)
	\biggr]
	+\frac{1}{72} \Bigl(
		16 \pi ^2 z^3-351 \zeta_3 z^2-48 \pi ^2 z^2-326 z^2+1314 \zeta_3 z+48 \pi ^2 z
	\nn\\&\quad						
		+172 z-702 \zeta_3+82 \pi^2+154
	\Bigr)
	+\frac{1}{6} \Bigl(-9 z^2+10 z+2\Bigr) \ln ^3(2)
	+\frac{1}{6} \Bigl(4 z^3-3 z^2+24 z
	\nn\\&\quad				
	+25\Bigr) \ln ^2(2)
	+\frac{1}{18} \Bigl(88 z^3-21 \pi ^2 z^2+216 z^2-24 \pi ^2 z+306 z-18 \pi ^2+178\Bigr)  \ln (2)
\Biggr\}
\nn\\&	
+
\frac{C_F}{z} \Biggl\{
	\frac{2}{3} (z-2) z \ln ^3(z)
	+\biggl[\frac{1}{2} (z-2) \ln (2) z +\frac{1}{4} \left(z^2+6 z+6\right)\biggr] \ln ^2(z)
\nn\\&		
	-\frac{1}{2} (z-2) z \ln \left(\frac{2 z}{1-z}\right) \ln ^2(z)
	+\biggl[
		\frac{1}{2} z \ln (2) (5 z+8)
		+\frac{1}{4} \left(-\pi ^2 z^2-16 z^2+2 \pi ^2 z-4 z+46\right)
	\nn\\&\quad					
		-\frac{1}{64} z (47 z-44)	\ln ^2(2)
	\biggr] \ln (z)
	-\frac{1}{6} (z-2) z \ln ^3\left(\frac{2 z}{z+1}\right)
	-\frac{1}{6} (z-2) z \ln ^3(z+1)
\nn\\&			
	+\biggl[\frac{1}{2} (z-2) \ln (z) z
	+\frac{1}{2}(z-2) \ln (2) z+\frac{1}{2} \left(2 z^2-z-3\right)\biggr] \ln ^2(z+1)
	+\frac{1}{24} \biggl[
		17 \pi ^2 z^2+162 z^2
	\nn\\&\quad		
		+6 \pi ^2 z-420 z-18 \pi^2+258
	\biggr]
	+\biggl[
		-\frac{1}{2} (z-2) z \ln ^2(z)
		+(z-2) z \ln \left(\frac{2 z}{1-z}\right) \ln (z)
	\nn\\&\quad				
		+\frac{1}{3} z \left(\pi ^2 z-2 \pi ^2\right)
		+\frac{5}{64} z (3	z+4) \ln ^2(2)
	\biggr] \ln \left(\frac{2 z}{z+1}\right)
	+ \ln (z+1)\biggl[
		-\frac{1}{64} z \ln ^2(2) (17 z-84)
	\nn\\&\quad		
		+\frac{1}{6} \left(2 \pi ^2 z^2+39 z^2-4 \pi ^2 z-15z-54\right)
		+\left(2 z^2-z-3\right) \ln (z)+\left(-2 z^2+z+3\right) \ln (2)
	\biggr]
\nn\\&	
	+\ln (1-z) \biggl[
		\ln (2) \left(2 z^2-z-3\right)
		+\frac{1}{12}\Bigl(-\pi ^2 z^2-30 z^2+2 \pi ^2 z+60 z-30\Bigr)			
		+\Bigl(-2 z^2
		\nn\\&\qquad				
			-(z-2) \ln (z) z+z+3\Bigr) \ln (z+1)
	\biggr]
	+\biggl[
		\frac{2}{3} z \ln ^3\left(\frac{2z}{z+1}\right)
		-z \ln \left(\frac{2 z}{1-z}\right) \ln ^2\left(\frac{2 z}{z+1}\right)
	\nn\\&\quad				
		-\frac{2}{3} \pi ^2 z \ln \left(\frac{2 z}{z+1}\right)
		+z \ln^3(z+1)
		-\frac{3}{4} z \ln ^2(z)
		+\left(-2 \ln (z) z-2 \ln (2) z+\frac{3 z}{4}\right) \ln ^2(z+1)
	\nn\\&\quad	
		-\frac{3 \pi ^2 z}{8}
		-\frac{1}{6} \left(27+\pi ^2\right) z \ln(z)
		+\biggl(
			-z \ln ^2(z)
			+\frac12 \Bigl(4 z \ln (2)-3 z \Bigr) \ln (z)
			-\frac{1}{6} \Bigl(-27+2 \pi ^2\Bigr) z
		\nn\\&\qquad				
			+z \ln ^2(2)
			-\frac{3}{2} z \ln (2)
		\biggr) \ln(z+1)
		+\ln ^2(1-z) \Bigl(-\ln (z) z-\ln (z+1) z+\ln (2) z \Bigr)
	\nn\\&\quad	
		+\ln (1-z) \left(
			-2 \ln (2) \ln (z) z-\frac{3}{2} \ln (2) z
			+\frac{\pi ^2 z}{2}
			+\Bigl(4 \ln (z) z+\frac{3	z}{2}\Bigr) \ln (z+1)
		\right)
	\nn\\&\quad	
		+\frac{3}{4} z \ln ^2(2)
		+\frac{1}{6} \left(-27+2 \pi ^2\right) z \ln (2)
	\biggr] P_{gq}(z)
	+\biggl[
		\frac{3}{2} \left(3 z^2+2z-2\right)
		-(z-2) z \ln (1-z)
	\nn\\&\quad		
		+(z-2) z \ln (z)
		+\left(-2 \ln \left(\frac{2 z}{1-z}\right) z+2 \ln (2) z-\frac{3 z}{2}\right) P_{gq}(z)
	\biggr]	\Li_2(-z)
	+\biggl[
		-2 z^2
	\nn\\&\quad		
		+\left(2 \ln \left(\frac{2 z}{1-z}\right) z-2 \ln (2) z+\frac{3 z}{2}\right) P_{gq}(z)+3
	\biggr]\Li_2(z)
	+\Li_2\left(\frac{z-1}{z+1}\right) \biggl[
		2z^2
		-z
	\nn\\&\quad		
		+\left(-2 \ln \left(\frac{2 z}{1-z}\right) z+2 \ln (2) z-\frac{3 z}{2}\right) P_{gq}(z)-3
	\biggr] 
	+(z-2) z\Li_3\left(\frac{1-z}{2}\right)
\nn\\&	
	-(z-2) z \Li_3(1-z)
	+\biggl[2 z P_{gq}(z)-(z-2) z \biggr] \Li_3\left(\frac{z-1}{2 z}\right)
	-\frac{1}{192} z \Bigl(13 z+124\Bigr) \ln^3(2)
\nn\\&		
	+\frac{1}{2} \Bigl(2 z^2-z-3\Bigr) \ln ^2(2)
	+\frac{1}{6} \Bigl(-2 \pi ^2 z^2-39 z^2+4 \pi ^2 z+15 z+54\Bigr) \ln (2)
\Biggr\}
\end{align}


\bibliographystyle{jhep}
\bibliography{beamfunc}

\providecommand{\href}[2]{#2}\begingroup\raggedright\begin{thebibliography}{10}

\bibitem{Stewart:2009yx}
I.~W. Stewart, F.~J. Tackmann and W.~J. Waalewijn, \emph{{Factorization at the
  LHC: From PDFs to Initial State Jets}},
  \href{http://dx.doi.org/10.1103/PhysRevD.81.094035}{\emph{Phys.~Rev.~D} {\bf
  81} (2010) 094035}, [\href{http://arxiv.org/abs/0910.0467}{{\tt 0910.0467}}].

\bibitem{Stewart:2010qs}
I.~W. Stewart, F.~J. Tackmann and W.~J. Waalewijn, \emph{{The Quark Beam
  Function at NNLL}},
  \href{http://dx.doi.org/10.1007/JHEP09(2010)005}{\emph{JHEP} {\bf 1009}
  (2010) 005}, [\href{http://arxiv.org/abs/1002.2213}{{\tt 1002.2213}}].

\bibitem{Jain:2011iu}
A.~Jain, M.~Procura and W.~J. Waalewijn, \emph{{Fully-Unintegrated Parton
  Distribution and Fragmentation Functions at Perturbative $k_T$}},
  \href{http://dx.doi.org/10.1007/JHEP04(2012)132}{\emph{JHEP} {\bf 1204}
  (2012) 132}, [\href{http://arxiv.org/abs/1110.0839}{{\tt 1110.0839}}].

\bibitem{Mantry:2009qz}
S.~Mantry and F.~Petriello, \emph{{Factorization and Resummation of Higgs Boson
  Differential Distributions in Soft-Collinear Effective Theory}},
  \href{http://dx.doi.org/10.1103/PhysRevD.81.093007}{\emph{Phys.~Rev.~D} {\bf
  81} (2010) 093007}, [\href{http://arxiv.org/abs/0911.4135}{{\tt 0911.4135}}].

\bibitem{Collins:2007ph}
J.~Collins, T.~Rogers and A.~Stasto, \emph{{Fully unintegrated parton
  correlation functions and factorization in lowest-order hard scattering}},
  \href{http://dx.doi.org/10.1103/PhysRevD.77.085009}{\emph{Phys.~Rev.~D} {\bf
  77} (2008) 085009}, [\href{http://arxiv.org/abs/0708.2833}{{\tt 0708.2833}}].

\bibitem{Rogers:2008jk}
T.~C. Rogers, \emph{{Next-to-Leading Order Hard Scattering Using Fully
  Unintegrated Parton Distribution Functions}},
  \href{http://dx.doi.org/10.1103/PhysRevD.78.074018}{\emph{Phys.~Rev.~D} {\bf
  78} (2008) 074018}, [\href{http://arxiv.org/abs/0807.2430}{{\tt 0807.2430}}].

\bibitem{Gaunt:2014xxa}
J.~R. Gaunt and M.~Stahlhofen, \emph{{The Fully-Differential Quark Beam
  Function at NNLO}},
  \href{http://dx.doi.org/10.1007/JHEP12(2014)146}{\emph{JHEP} {\bf 12} (2014)
  146}, [\href{http://arxiv.org/abs/1409.8281}{{\tt 1409.8281}}].

\bibitem{Mulders:2000sh}
P.~J. Mulders and J.~Rodrigues, \emph{{Transverse momentum dependence in gluon
  distribution and fragmentation functions}},
  \href{http://dx.doi.org/10.1103/PhysRevD.63.094021}{\emph{Phys. Rev.} {\bf
  D63} (2001) 094021}, [\href{http://arxiv.org/abs/hep-ph/0009343}{{\tt
  hep-ph/0009343}}].

\bibitem{Tackmann:2012bt}
F.~J. Tackmann, J.~R. Walsh and S.~Zuberi, \emph{{Resummation Properties of Jet
  Vetoes at the LHC}},
  \href{http://dx.doi.org/10.1103/PhysRevD.86.053011}{\emph{Phys.~Rev.~D} {\bf
  86} (2012) 053011}, [\href{http://arxiv.org/abs/1206.4312}{{\tt 1206.4312}}].

\bibitem{Procura:2014cba}
M.~Procura, W.~J. Waalewijn and L.~Zeune, \emph{{Resummation of
  Double-Differential Cross Sections and Fully-Unintegrated Parton Distribution
  Functions}}, \href{http://dx.doi.org/10.1007/JHEP02(2015)117}{\emph{JHEP}
  {\bf 02} (2015) 117}, [\href{http://arxiv.org/abs/1410.6483}{{\tt
  1410.6483}}].

\bibitem{Gangal:2014qda}
S.~Gangal, M.~Stahlhofen and F.~J. Tackmann, \emph{{Rapidity-Dependent Jet
  Vetoes}}, \href{http://dx.doi.org/10.1103/PhysRevD.91.054023}{\emph{Phys.
  Rev. D} {\bf 91} (2015) 054023}, [\href{http://arxiv.org/abs/1412.4792}{{\tt
  1412.4792}}].

\bibitem{Aaboud:2018jqu}
{\scshape ATLAS} collaboration, M.~Aaboud et~al., \emph{{Measurements of
  gluon-gluon fusion and vector-boson fusion Higgs boson production
  cross-sections in the $H \to WW^{\ast} \to e\nu\mu\nu$ decay channel in $pp$
  collisions at $\sqrt{s}=13$ TeV with the ATLAS detector}},
  \href{http://dx.doi.org/10.1016/j.physletb.2018.11.064}{\emph{Phys. Lett.}
  {\bf B789} (2019) 508--529}, [\href{http://arxiv.org/abs/1808.09054}{{\tt
  1808.09054}}].

\bibitem{Sirunyan:2018egh}
{\scshape CMS} collaboration, A.~M. Sirunyan et~al., \emph{{Measurements of
  properties of the Higgs boson decaying to a W boson pair in pp collisions at
  $\sqrt{s}=$ 13 TeV}},
  \href{http://dx.doi.org/10.1016/j.physletb.2018.12.073}{\emph{Phys. Lett.}
  {\bf B791} (2019) 96}, [\href{http://arxiv.org/abs/1806.05246}{{\tt
  1806.05246}}].

\bibitem{Aaboud:2018xdt}
{\scshape ATLAS} collaboration, M.~Aaboud et~al., \emph{{Measurements of Higgs
  boson properties in the diphoton decay channel with 36 fb$^{-1}$ of $pp$
  collision data at $\sqrt{s} = 13$ TeV with the ATLAS detector}},
  \href{http://dx.doi.org/10.1103/PhysRevD.98.052005}{\emph{Phys. Rev.} {\bf
  D98} (2018) 052005}, [\href{http://arxiv.org/abs/1802.04146}{{\tt
  1802.04146}}].

\bibitem{Alioli:2012fc}
S.~Alioli, C.~W. Bauer, C.~J. Berggren, A.~Hornig, F.~J. Tackmann et~al.,
  \emph{{Combining Higher-Order Resummation with Multiple NLO Calculations and
  Parton Showers in GENEVA}},
  \href{http://dx.doi.org/10.1007/JHEP09(2013)120}{\emph{JHEP} {\bf 1309}
  (2013) 120}, [\href{http://arxiv.org/abs/1211.7049}{{\tt 1211.7049}}].

\bibitem{Alioli:2013hqa}
S.~Alioli, C.~W. Bauer, C.~Berggren, F.~J. Tackmann, J.~R. Walsh et~al.,
  \emph{{Matching Fully Differential NNLO Calculations and Parton Showers}},
  \href{http://dx.doi.org/10.1007/JHEP06(2014)089}{\emph{JHEP} {\bf 1406}
  (2014) 089}, [\href{http://arxiv.org/abs/1311.0286}{{\tt 1311.0286}}].

\bibitem{Alioli:2015toa}
S.~Alioli, C.~W. Bauer, C.~Berggren, F.~J. Tackmann and J.~R. Walsh,
  \emph{{Drell-Yan production at NNLL$'+$NNLO matched to parton showers}},
  \href{http://dx.doi.org/10.1103/PhysRevD.92.094020}{\emph{Phys. Rev.} {\bf
  D92} (2015) 094020}, [\href{http://arxiv.org/abs/1508.01475}{{\tt
  1508.01475}}].

\bibitem{Alioli:2019qzz}
S.~Alioli, A.~Broggio, S.~Kallweit, M.~A. Lim and L.~Rottoli,
  \emph{{Higgsstrahlung at NNLL$'+$NNLO matched to parton showers in GENEVA}},
  \href{http://dx.doi.org/10.1103/PhysRevD.100.096016}{\emph{Phys. Rev.} {\bf
  D100} (2019) 096016}, [\href{http://arxiv.org/abs/1909.02026}{{\tt
  1909.02026}}].

\bibitem{Gaunt:2015pea}
J.~Gaunt, M.~Stahlhofen, F.~J. Tackmann and J.~R. Walsh, \emph{{N-jettiness
  Subtractions for NNLO QCD Calculations}},
  \href{http://dx.doi.org/10.1007/JHEP09(2015)058}{\emph{JHEP} {\bf 09} (2015)
  058}, [\href{http://arxiv.org/abs/1505.04794}{{\tt 1505.04794}}].

\bibitem{Boughezal:2015dva}
R.~Boughezal, C.~Focke, X.~Liu and F.~Petriello, \emph{{$W$-boson production in
  association with a jet at next-to-next-to-leading order in perturbative
  QCD}}, \href{http://dx.doi.org/10.1103/PhysRevLett.115.062002}{\emph{Phys.\
  Rev.\ Lett.} {\bf 115} (2015) 062002},
  [\href{http://arxiv.org/abs/1504.02131}{{\tt 1504.02131}}].

\bibitem{Lustermans:2019cau}
G.~Lustermans, J.~K.~L. Michel and F.~J. Tackmann, \emph{{Generalized Threshold
  Factorization with Full Collinear Dynamics}},
  \href{http://arxiv.org/abs/1908.00985}{{\tt 1908.00985}}.

\bibitem{Gaunt:2014xga}
J.~R. Gaunt, M.~Stahlhofen and F.~J. Tackmann, \emph{{The Quark Beam Function
  at Two Loops}}, \href{http://dx.doi.org/10.1007/JHEP04(2014)113}{\emph{JHEP}
  {\bf 04} (2014) 113}, [\href{http://arxiv.org/abs/1401.5478}{{\tt
  1401.5478}}].

\bibitem{Gaunt:2014cfa}
J.~Gaunt, M.~Stahlhofen and F.~J. Tackmann, \emph{{The Gluon Beam Function at
  Two Loops}}, \href{http://dx.doi.org/10.1007/JHEP08(2014)020}{\emph{JHEP}
  {\bf 1408} (2014) 020}, [\href{http://arxiv.org/abs/1405.1044}{{\tt
  1405.1044}}].

\bibitem{Bauer:2000ew}
C.~W. Bauer, S.~Fleming and M.~E. Luke, \emph{{Summing Sudakov logarithms in
  $B\to X_s\gamma$ in effective field theory}},
  \href{http://dx.doi.org/10.1103/PhysRevD.63.014006}{\emph{Phys.~Rev.~D} {\bf
  63} (2000) 014006}, [\href{http://arxiv.org/abs/hep-ph/0005275}{{\tt
  hep-ph/0005275}}].

\bibitem{Bauer:2000yr}
C.~W. Bauer, S.~Fleming, D.~Pirjol and I.~W. Stewart, \emph{{An Effective field
  theory for collinear and soft gluons: Heavy to light decays}},
  \href{http://dx.doi.org/10.1103/PhysRevD.63.114020}{\emph{Phys.~Rev.~D} {\bf
  63} (2001) 114020}, [\href{http://arxiv.org/abs/hep-ph/0011336}{{\tt
  hep-ph/0011336}}].

\bibitem{Bauer:2001ct}
C.~W. Bauer and I.~W. Stewart, \emph{{Invariant operators in collinear
  effective theory}},
  \href{http://dx.doi.org/10.1016/S0370-2693(01)00902-9}{\emph{Phys.~Lett.~B}
  {\bf 516} (2001) 134--142}, [\href{http://arxiv.org/abs/hep-ph/0107001}{{\tt
  hep-ph/0107001}}].

\bibitem{Bauer:2001yt}
C.~W. Bauer, D.~Pirjol and I.~W. Stewart, \emph{{Soft collinear factorization
  in effective field theory}},
  \href{http://dx.doi.org/10.1103/PhysRevD.65.054022}{\emph{Phys.~Rev.~D} {\bf
  65} (2002) 054022}, [\href{http://arxiv.org/abs/hep-ph/0109045}{{\tt
  hep-ph/0109045}}].

\bibitem{Bauer:2002nz}
C.~W. Bauer, S.~Fleming, D.~Pirjol, I.~Z. Rothstein and I.~W. Stewart,
  \emph{{Hard scattering factorization from effective field theory}},
  \href{http://dx.doi.org/10.1103/PhysRevD.66.014017}{\emph{Phys.~Rev.~D} {\bf
  66} (2002) 014017}, [\href{http://arxiv.org/abs/hep-ph/0202088}{{\tt
  hep-ph/0202088}}].

\bibitem{Beneke:2002ph}
M.~Beneke, A.~Chapovsky, M.~Diehl and T.~Feldmann, \emph{{Soft collinear
  effective theory and heavy to light currents beyond leading power}},
  \href{http://dx.doi.org/10.1016/S0550-3213(02)00687-9}{\emph{Nucl.~Phys.}
  {\bf B643} (2002) 431--476}, [\href{http://arxiv.org/abs/hep-ph/0206152}{{\tt
  hep-ph/0206152}}].

\bibitem{Luebbert:2016itl}
T.~Lübbert, J.~Oredsson and M.~Stahlhofen, \emph{{Rapidity renormalized TMD
  soft and beam functions at two loops}},
  \href{http://dx.doi.org/10.1007/JHEP03(2016)168}{\emph{JHEP} {\bf 03} (2016)
  168}, [\href{http://arxiv.org/abs/1602.01829}{{\tt 1602.01829}}].

\bibitem{Zhu:2012ts}
H.~X. Zhu, C.~S. Li, H.~T. Li, D.~Y. Shao and L.~L. Yang,
  \emph{{Transverse-momentum resummation for top-quark pairs at hadron
  colliders}},
  \href{http://dx.doi.org/10.1103/PhysRevLett.110.082001}{\emph{Phys. Rev.
  Lett.} {\bf 110} (2013) 082001}, [\href{http://arxiv.org/abs/1208.5774}{{\tt
  1208.5774}}].

\bibitem{Li:2013mia}
H.~T. Li, C.~S. Li, D.~Y. Shao, L.~L. Yang and H.~X. Zhu, \emph{{Top quark pair
  production at small transverse momentum in hadronic collisions}},
  \href{http://dx.doi.org/10.1103/PhysRevD.88.074004}{\emph{Phys. Rev.} {\bf
  D88} (2013) 074004}, [\href{http://arxiv.org/abs/1307.2464}{{\tt
  1307.2464}}].

\bibitem{Fleming:2006cd}
S.~Fleming, A.~K. Leibovich and T.~Mehen, \emph{{Resummation of Large Endpoint
  Corrections to Color-Octet $J/\psi$ Photoproduction}},
  \href{http://dx.doi.org/10.1103/PhysRevD.74.114004}{\emph{Phys.~Rev.~D} {\bf
  74} (2006) 114004}, [\href{http://arxiv.org/abs/hep-ph/0607121}{{\tt
  hep-ph/0607121}}].

\bibitem{Berger:2010xi}
C.~F. Berger, C.~Marcantonini, I.~W. Stewart, F.~J. Tackmann and W.~J.
  Waalewijn, \emph{{Higgs Production with a Central Jet Veto at NNLL+NNLO}},
  \href{http://dx.doi.org/10.1007/JHEP04(2011)092}{\emph{JHEP} {\bf 1104}
  (2011) 092}, [\href{http://arxiv.org/abs/1012.4480}{{\tt 1012.4480}}].

\bibitem{Moch:2004pa}
S.~Moch, J.~Vermaseren and A.~Vogt, \emph{{The Three loop splitting functions
  in QCD: The Nonsinglet case}},
  \href{http://dx.doi.org/10.1016/j.nuclphysb.2004.03.030}{\emph{Nucl.~Phys.}
  {\bf B688} (2004) 101--134}, [\href{http://arxiv.org/abs/hep-ph/0403192}{{\tt
  hep-ph/0403192}}].

\bibitem{Vogt:2004mw}
A.~Vogt, S.~Moch and J.~Vermaseren, \emph{{The Three-loop splitting functions
  in QCD: The Singlet case}},
  \href{http://dx.doi.org/10.1016/j.nuclphysb.2004.04.024}{\emph{Nucl.~Phys.}
  {\bf B691} (2004) 129--181}, [\href{http://arxiv.org/abs/hep-ph/0404111}{{\tt
  hep-ph/0404111}}].

\bibitem{Becher:2009th}
T.~Becher and M.~D. Schwartz, \emph{{Direct photon production with effective
  field theory}}, \href{http://dx.doi.org/10.1007/JHEP02(2010)040}{\emph{JHEP}
  {\bf 02} (2010) 040}, [\href{http://arxiv.org/abs/0911.0681}{{\tt
  0911.0681}}].

\bibitem{Moch:2018wjh}
S.~Moch, B.~Ruijl, T.~Ueda, J.~M. Vermaseren and A.~Vogt, \emph{{On quartic
  colour factors in splitting functions and the gluon cusp anomalous
  dimension}},
  \href{http://dx.doi.org/10.1016/j.physletb.2018.06.017}{\emph{Phys.\ Lett.\
  B} {\bf 782} (2018) 627--632}, [\href{http://arxiv.org/abs/1805.09638}{{\tt
  1805.09638}}].

\bibitem{Grozin:2018vdn}
A.~Grozin, \emph{{Four-loop cusp anomalous dimension in QED}},
  \href{http://dx.doi.org/10.1007/JHEP01(2019)134}{\emph{JHEP} {\bf 06} (2018)
  073}, [\href{http://arxiv.org/abs/1805.05050}{{\tt 1805.05050}}].

\bibitem{Lee:2019zop}
R.~N. Lee, A.~V. Smirnov, V.~A. Smirnov and M.~Steinhauser, \emph{{Four-loop
  quark form factor with quartic fundamental colour factor}},
  \href{http://dx.doi.org/10.1007/JHEP02(2019)172}{\emph{JHEP} {\bf 02} (2019)
  172}, [\href{http://arxiv.org/abs/1901.02898}{{\tt 1901.02898}}].

\bibitem{Henn:2019rmi}
J.~Henn, T.~Peraro, M.~Stahlhofen and P.~Wasser, \emph{{Matter dependence of
  the four-loop cusp anomalous dimension}},
  \href{http://dx.doi.org/10.1103/PhysRevLett.122.201602}{\emph{Phys.\ Rev.\
  Lett.} {\bf 122} (2019) 201602}, [\href{http://arxiv.org/abs/1901.03693}{{\tt
  1901.03693}}].

\bibitem{Bruser:2019auj}
R.~Brüser, A.~Grozin, J.~M. Henn and M.~Stahlhofen, \emph{{Matter dependence
  of the four-loop QCD cusp anomalous dimension: from small angles to all
  angles}}, \href{http://dx.doi.org/10.1007/JHEP05(2019)186}{\emph{JHEP} {\bf
  05} (2019) 186}, [\href{http://arxiv.org/abs/1902.05076}{{\tt 1902.05076}}].

\bibitem{Henn:2019swt}
J.~M. Henn, G.~P. Korchemsky and B.~Mistlberger, \emph{{The full four-loop cusp
  anomalous dimension in $\mathcal{N}=4$ super Yang-Mills and QCD}},
  \href{http://arxiv.org/abs/1911.10174}{{\tt 1911.10174}}.

\bibitem{vonManteuffel:2020vjv}
A.~von Manteuffel, E.~Panzer and R.~M. Schabinger, \emph{{Analytic four-loop
  anomalous dimensions in massless QCD from form factors}},
  \href{http://arxiv.org/abs/2002.04617}{{\tt 2002.04617}}.

\end{thebibliography}\endgroup

\end{document}